%% file: Alahmed_Tong_ECstorage_v3.tex
\documentclass[journal]{IEEEtran}
\usepackage[left=0.75in, right=0.75in, top=1in, bottom=0.9in]{geometry}
\makeatletter
\def\ps@headings{%
\def\@oddhead{\mbox{}\scriptsize\rightmark \hfil \thepage}%
\def\@evenhead{\scriptsize\thepage \hfil \leftmark\mbox{}}%
\def\@oddfoot{}%
\def\@evenfoot{}}
\makeatother 

\usepackage{amsmath,amssymb,mathrsfs,amsthm,bm}
\usepackage{mathtools}
\usepackage{cite}

\usepackage{graphicx,stfloats,subfigure,multicol}
\usepackage{epsfig,epsf,psfrag,amssymb,amsfonts,latexsym,graphicx,mathrsfs,subfigure}
\usepackage[table,xcdraw]{xcolor}
\usepackage{lettrine}
\usepackage{comment}
\usepackage{enumitem}
\newlist{enumsteps}{enumerate}{2}
\setlist[enumsteps,1]{label=Case \arabic*: }
\setlist[enumsteps,2]{label=Case \arabic{enumstepsi}.\arabic*: }
\usepackage{booktabs}
\usepackage[table]{xcolor}
\usepackage{multirow}
\usepackage{tablefootnote}
\usepackage{bbm}
\usepackage{orcidlink}
\usepackage{chngpage}
\usepackage{textcomp}
\usepackage{hyperref}
\usepackage{pifont}
\usepackage{url}
\usepackage[hyphenbreaks]{breakurl}
\usepackage[normalem]{ulem}
\usepackage{algpseudocode}
\newsavebox{\ieeealgbox}

\usepackage{array}

\newtheorem{theorem}{Theorem}
\newtheorem{proposition}{Proposition}
\newtheorem{corollary}{Corollary}
\newtheorem{lemma}{Lemma}
\newtheorem{definition}{Definition}

\newtheorem{axiom}{Axiom}
\newtheorem*{policy*}{Dynamic NEM}
\newtheorem*{policy1*}{Generalized Dynamic NEM}

 \def\old#1{}

\input LTmacros

\begin{document}

\title{Dynamic Net Metering for Energy Communities
}

\author{Ahmed S. Alahmed\orcidlink{0000-0002-4715-4379},~\IEEEmembership{Graduate~Student~Member,~IEEE} and
Lang~Tong\orcidlink{0000-0003-3322-2681},~\IEEEmembership{Fellow,~IEEE}
\thanks{\scriptsize  Ahmed S. Alahmed and
Lang Tong ({\tt \{\tcb{ASA278,~LT35}\}\tcb{@cornell.edu}}) are  with the School of Electrical and Computer Engineering, Cornell University, USA.}
\thanks{\scriptsize This work was supported in part by the National Science Foundation under
Award 2218110 and the Power Systems and Engineering Research Center (PSERC) Research Project M-46.}
}

\maketitle
\begin{abstract}
We propose a social welfare maximizing market mechanism for an energy community that aggregates individual and community-shared energy resources under a general net energy metering (NEM) policy. Referred to as Dynamic NEM (D-NEM), the proposed mechanism dynamically sets the community NEM prices based on aggregated community resources, including flexible consumption, storage, and renewable generation. D-NEM guarantees a higher benefit to each community member than possible outside the community, and no sub-communities would be better off departing from its parent community. D-NEM aligns each member's incentive with that of the community such that each member maximizing individual surplus under D-NEM results in maximum community social welfare. Empirical studies compare the proposed mechanism with existing benchmarks, demonstrating its welfare benefits, operational characteristics, and responsiveness to NEM rates.
\end{abstract}


\section{Introduction}\label{sec:intro}
\input{intro_v2.tex}

\section{Energy Community Model}\label{sec:form}
\input{formulation_v2.tex}

\section{Decentralized Welfare Maximization}\label{sec:DECWel}
\input{DECWel_v2}

\section{Stackelberg Equilibrium Policy and Properties}\label{sec:DNEM1}
\input{DNEM1_v2.tex}
\section{Dynamic NEM With BESS}\label{sec:DNEM2}
\input{DNEM2_v2.tex}

\section{Numerical Results}\label{sec:num}
\input{num_v1.tex}

\section{Conclusion}\label{sec:conclusion}
  This paper proposed Dynamic NEM (D-NEM)--- a social welfare maximizing market mechanism for energy communities that aggregates BTM and community-shared DER under a general NEM policy. Through D-NEM, the community's flexible demands and BESS, become DG-aware by being dynamically scheduled based on the aggregate renewables. We showed that D-NEM and individual surplus maximization is a Nash equilibrium that ensures higher surplus for community members than the maximum surplus under DSO’s NEM. We also showed that the maximum community welfare is achieved under the equilibrium solution. We proved that D-NEM satisfies the cost-causation principle and stabilize the coalition by ensuring that no sub-coalitions would find it advantageous to leave the grand coalition.

An interesting future direction is to explore the notion of {\em bounded rationality}, under which prosumers may not be surplus-maximizers, perhaps due to the absence of HEMS. Furthermore, we ignored in this work the long-term planning study which allows us to address issues like central BESS and solar sizing, capital costs of community future investments, and the evolution of community member size and shares of the central BESS.

\section*{Acknowledgment}
The authors are grateful for the insights and discussions with Prof. Timothy D. Mount from Cornell University.
{
\bibliographystyle{IEEEtran}
\bibliography{EnergyComm}
}
\section*{Appendix A: Dynamic NEM Numerical Example}\label{sec:NumExample}
\input AppendixExample
\section*{Appendix B: Framework Generalization}\label{sec:FrameworkGeneral}
\input AppendixGeneralization
\section*{Appendix C: Proofs}\label{sec:appendix_proofs}
\input{AppProofs/ProofsMain_v2}
\section*{Appendix D: Generalized Dynamic NEM}\label{sec:appendix_GenDNEM}
\input AppendixGenDNEM

\begin{IEEEbiography}[{\includegraphics[width=0.98in,height=1.20in,clip]{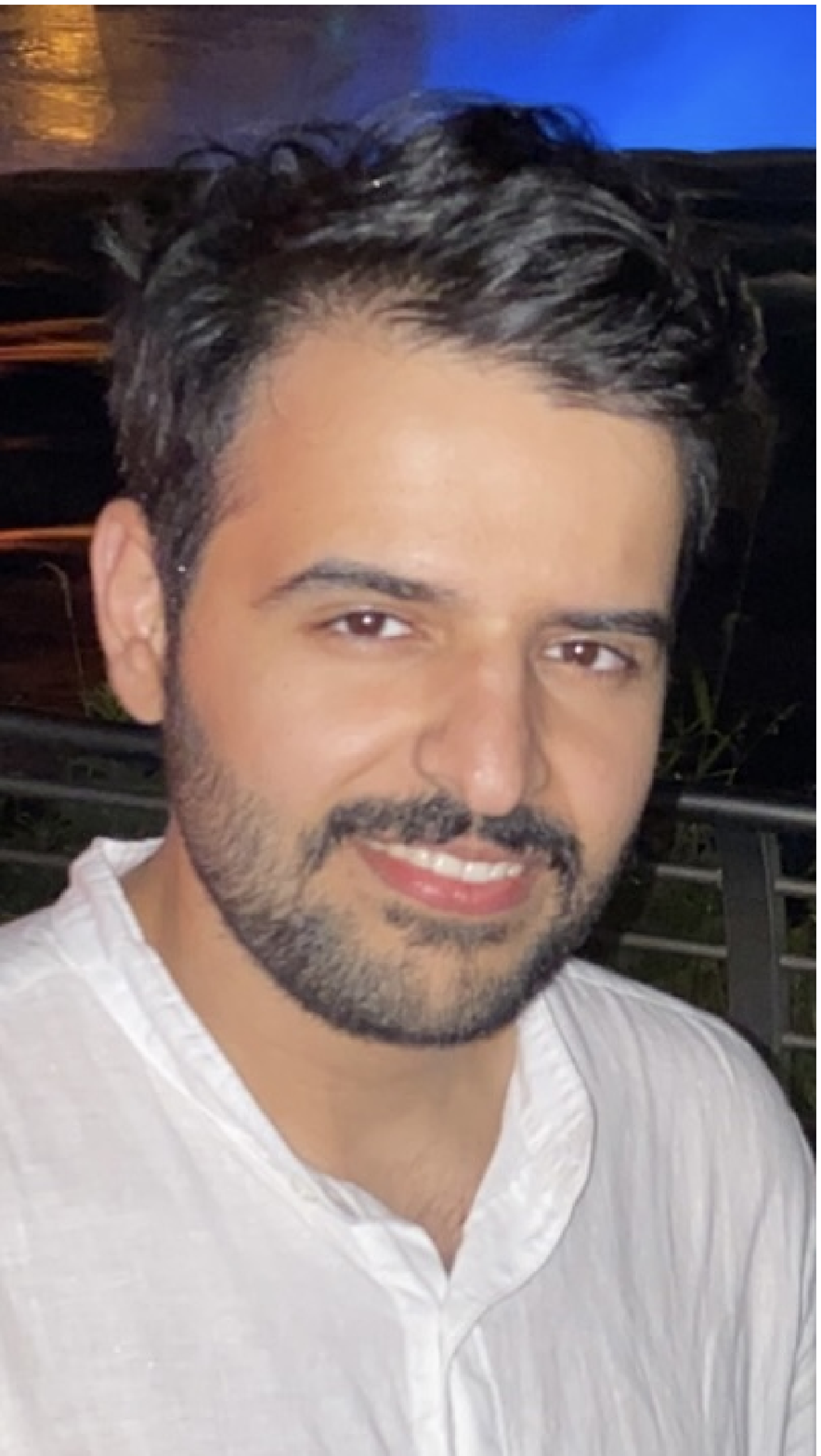}}]{Ahmed S. Alahmed}(S'14) is an Electrical and Computer Engineering (ECE) Ph.D. candidate at Cornell University. He received his B.Sc. and M.Sc. in ECE from King Fahd University of Petroleum and Minerals in 2016 and 2019, respectively. Ahmed has worked as a Ph.D. research intern at the National Renewable Energy Lab (NREL), the Electric Power Research Institute (EPRI), and The Brattle Group.  In addition to receiving two best paper awards, he received, in 2022, the esteemed MiSK Fellowship Award for empowering future Saudi leaders.
\par Ahmed's research interests include energy systems control and optimization, DER aggregation and adoption dynamics, game theory, and mechanisms design for transactive energy systems.
\end{IEEEbiography}
\vspace{-12cm}
\begin{IEEEbiography}[{\includegraphics[width=1.5in,height=1.25in,clip,keepaspectratio]{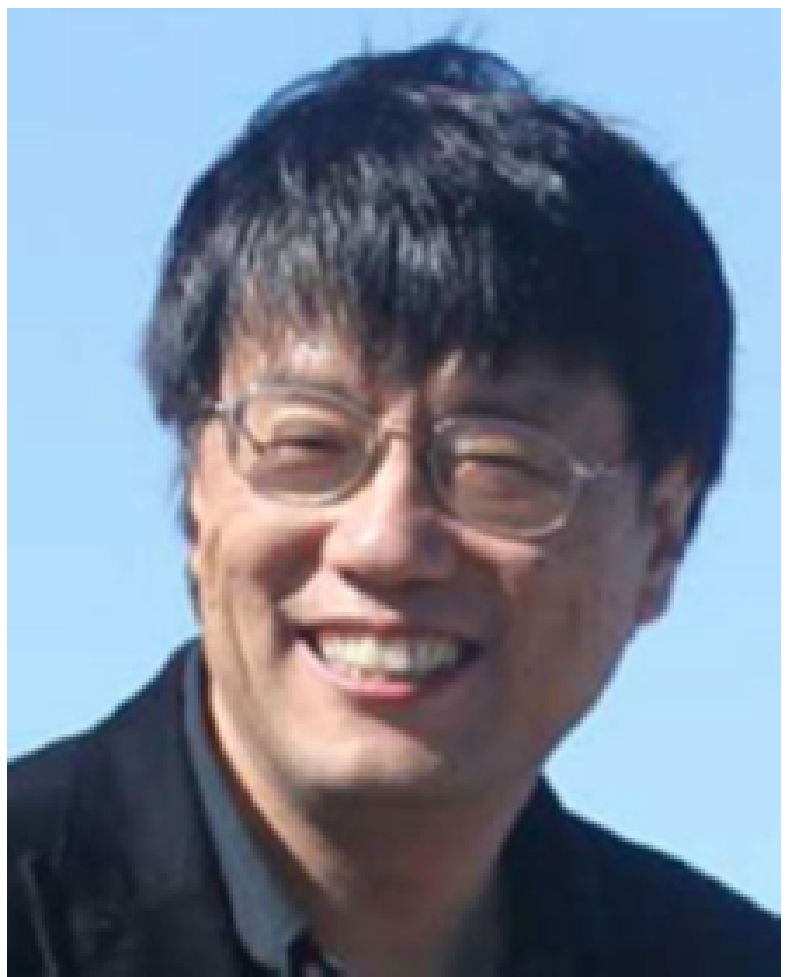}}]{Lang Tong}(S'87, M'91, SM'01, F'05) is the Irwin and Joan Jacobs Professor in Engineering and the Cornell Site Director of the Power Systems Engineering Research Center (PSERC). His current research focuses on energy and power systems, electricity markets, smart grids, and the electrification of transportation systems. His expertise lies in data analytics, machine learning, optimization, and market design. He received numerous publication awards from the IEEE Signal Processing, Communications, and PES Societies.
 
Lang Tong received a B.E. degree from Tsinghua University and a Ph.D. degree from the University of Notre Dame. He was a Postdoctoral Research Affiliate at Stanford University and held visiting positions at Stanford University, the University of California at Berkeley, the Delft University of Technology, and the Chalmers University of Technology in Sweden.   He was the 2018 Fulbright Distinguished Chair in Alternative Energy.
\end{IEEEbiography}
\end{document}

%% file: LTmacros.tex
\def\nn{\nonumber}
\def\beq{\begin{equation}}
\def\eeq{\end{equation}}
\def\bea{\begin{eqnarray}}
\def\eea{\end{eqnarray}}
\def\ba{\begin{array}}
\def\ea{\end{array}}

\def\bitem{\begin{itemize}}
\def\eitem{\end{itemize}}
\def\ben{\begin{enumerate}}
\def\een{\end{enumerate}}

\def\ie{{\it i.e.,\ \/}}

\definecolor{bgrd}{rgb}{1,1,1}
\definecolor{gray}{rgb}{0.5,0.5,0.5}
\definecolor{dkr}{rgb}{0.7,0.1,0.2}
\definecolor{dkb}{rgb}{0.1,0.1,0.8}

\def\tcb{\textcolor{blue}}

\newcommand{\mbbE}{\mathbb{E}}

\newcommand{\Xmsc}{\mathscr{X}}

\def\gbf{{\bf g}}

\def\Ac{{\cal A}}
\def\Bc{{\cal B}}

\def\Hc{{\cal H}}

\def\Kc{{\cal K}}

\def\Nc{{\cal N}}

\def\Sc{{\cal S}}
\def\Tc{{\cal T}}

%% file: intro_v2.tex
\lettrine{W}{e consider} the problem of pricing of electricity within a self-organized {\em energy community} (a.k.a. energy coalition) in a distribution system operated by a regulated distribution system operator\footnote{We, interchangeably, use the terms distribution utility and DSO.} (DSO).  Generically, energy communities consist of consumers and prosumers with behind-the-meter (BTM) and possibly shared distributed energy resources (DERs) such as solar, wind, and battery storage \cite{parag&sovacool:16Nature,Lezama&Soares&Hernandez:19TPS}. Examples of such an energy community include some of the closed distribution systems such as citizen energy communities \cite{CitizenEC:EuroCommWebsite}, community choice aggregators (CCAs), housing cooperatives, and educational/medical campuses, where the pricing policy of electricity within the community is set by its members through an energy community operator.  In this work, we assume that the aggregated energy consumption and production are subject to the regulated NEM tariff imposed by a DSO.  In particular, the DSO measures the community’s net consumption and charges the community at a {\em buy (import) rate} if the community is net-importing and a {\em sell (export) rate} if net-exporting.


The {\em sharing economy} concept has disrupted the transportation and accommodation markets by enabling the sharing of idle or underutilized resources \cite{Cohen&Kietzmann:14OrgEnvJSTOR}. Ascribed as a manifestation of the {\em sharing economy} concept, energy communities facilitate resource sharing between owners and users, which maximizes DER utilization and harnesses DER flexibility. Energy communities have gained tremendous traction recently due to enabling programs and rulings in addition to their role in improving system efficiency and economies of scale. Several public utility commissions have initiated energy-community-enabling programs, NEM aggregation (NEMA) and virtual NEM (VNEM) programs,\footnote{See, for example, Pacific Gas and Electric company (\href{https://www.pge.com/en_US/residential/solar-and-vehicles/options/option-overview/how-to-get-started/nema/net-energy-metering-aggregation.page}{PG\&E}), California, and Baltimore Gas and Electric Company (\href{https://www.bge.com/MyAccount/MyBillUsage/Documents/Electric/Rdr_18.pdf}{BGE}), Maryland.} for university campuses, residential complexes, and medical cities\cite{NEMevolution:23NAS}. In Europe, the majority of EU countries passed rules allowing standalone members and apartment complexes to form energy communities and developed jurisdictions to regulate and govern the operation of such coalitions \cite{Abada&Ehrenmann&Lambin:20EP}.

Several challenges, however, still face the wider adoption of energy communities, including financial barriers, such as the high initial costs of renewable energy and storage installations, grid connection, and metering expenses. A major challenge is how can energy communities compete with the regulated DSO who offers lucrative NEM incentives to its customers, which invites analyzing the design of a community pricing mechanism that is just and fair to all community members, and competitive to the DSO's tariff.

We focus in this work on the design of a DER-aware community market mechanism that (i) aligns individual incentives with the community welfare-maximizing objective and (ii) provides higher benefit for every subset of community members than it would have achieved if they were outside the community, thus ensuring the stability of the community. 


\subsection{Related Work}
There is a rich literature on energy communities covering cost-sharing mechanisms \cite{Yang&Guoqiang&Spanos:21TSG,Chakraborty&Poolla&Varaiya:19TSG,Chis&Koivunen:19TSG}, optimal energy management \cite{Han&Morstyn&McCulloch:19TPS,Fleischhacker&Auer&Lettner&Botterud:19TSG,Cui&Wang&Yan&Shi&Xiao:21TSG}, and coordination frameworks \cite{Guerreroetal:20RSER,Hao&Wu&Lian&Yang:188TSG,Kalathil&Wu&Poolla&Varaiya:19TSG}. Most relevant to this work are the community pricing and allocation rules \cite{Chakraborty&Poolla&Varaiya:19TSG,Yang&Guoqiang&Spanos:21TSG}, 
and welfare maximizing resource allocation \cite{Fleischhacker&Auer&Lettner&Botterud:19TSG,Cui&Wang&Yan&Shi&Xiao:21TSG}. A holistic review of energy community market mechanisms can be found in \cite{Tsaousoglou&Giraldo&Paterakis:22RSER}.

Three energy community models have been widely discussed, each offering a different market hierarchy and flexibility to its members. The first is the decentralized peer-to-peer (P2P) model \cite{Morstyn&Teytelboym&Mcculloch:19TSG,Sorin&Bobo&Pinson:19TPS}. Through {\em bilateral contracts}, the P2P market structure gives full flexibility and privacy to its members and the benefits of robustness in decentralization. The P2P market structure, however, is often challenged by policy and physical restrictions of a practical power system, accurate auditing of power delivery, and convergence to social optimality.

The second is the centralized model involving a community operator scheduling all resources for the benefit of the community \cite{Prete&Hobbs:16AE,Han&Morstyn&McCulloch:19TPS, Chis&Koivunen:19TSG}. While this model has the potential to achieve the highest community welfare, it often comes with substantial computation costs and privacy concerns. In particular, optimal central scheduling typically requires the operator to know the consumption preferences of individual members such as their time-of-use patterns and consumption priorities, for which an individual may be reluctant to share for privacy reasons.  A significant issue is also that optimizing community welfare may require trading off the benefits of some members with others.  In other words, maximizing community welfare may not align with individual surplus maximization.

The third model, to which the work presented here belongs, is the decentralized scheduling of community resources through a pricing mechanism \cite{Yang&Guoqiang&Spanos:21TSG,Chakraborty&Poolla&Varaiya:19TSG,Cui&Wang&Li&Xiao:20,Fleischhacker&Auer&Lettner&Botterud:19TSG,Cui&Wang&Yan&Shi&Xiao:21TSG,Celik&Roche&Bouquain&Miraoui:18TSG,Chakraborty&Baeyens&Poolla&Khargonekar&Varaiya:19TSG}. In \cite{Fleischhacker&Auer&Lettner&Botterud:19TSG}, a bi-level optimization of an apartment building energy community with DER owners is formulated with pricing and cost allocation mechanisms. It is unclear whether such a model will achieve individual and group rationality, without which the community coalition is not stable. In \cite{Cui&Wang&Yan&Shi&Xiao:21TSG}, a Nash bargaining benefit-sharing model is proposed to ensure that the community members will not abandon the community. A well-known disadvantage of such cost re-distribution schemes is that the computation complexity of cost allocation grows exponentially with the size of the community. Also, as in \cite{Fleischhacker&Auer&Lettner&Botterud:19TSG}, the competitiveness to lucrative retail programs such as NEM is not addressed. A low computation allocation rule is proposed in \cite{Chakraborty&Baeyens&Poolla&Khargonekar&Varaiya:19TSG} for sharing the realized cost of an energy community with storage and inflexible consumption under the time-of-use (ToU) rates. However, the bi-directionality of power flow, which is the essence of NEM policies, is not considered, as customers were assumed to be always net-importing, and therefore one cannot argue the competitiveness to NEM. The authors in \cite{Cui&Wang&Li&Xiao:20} analyze a stochastic energy community model with cost-minimizing members. An algorithm is proposed for better estimation of the stochastic game. The authors, however, assume equal energy buy and sell rates, which significantly limits the model, and the optimal resource scheduling because the BTM DG output provides the same value regardless of whether it was self-consumed or exported back to the grid. The work in \cite{Celik&Roche&Bouquain&Miraoui:18TSG} proposes a decentralized approach to energy management via flexible loads in energy communities. It is shown that community members solve their cost minimization problem using a genetic algorithm, after which the community operator aggregates the net consumptions and announces the updated community prices, iterating this process until convergence. Lastly, cost allocation and optimal operation and sizing of shared energy community storage are studied in \cite{Yang&Guoqiang&Spanos:21TSG}. The work, however, does not consider flexible demands, and the proposed {\em ex-post} cost allocation requires an algorithm to search for the nucleolus of a coalition game.

The work of Chakraborty et al. \cite{Chakraborty&Poolla&Varaiya:19TSG} stands out as the first mechanism design under the cost-causation principle, which offers every community member a lower payment than would be outside the community when the customer faces NEM. Unlike nucleolus, Shapely value, and Nash bargaining solutions \cite{Han&Morstyn&McCulloch:19TPS,Shapley&Shubik:73RAND,Cui&Wang&Li&Xiao:20}, among others, the computation complexity of the payment rule in \cite{Chakraborty&Poolla&Varaiya:19TSG} does not increase with the cardinality of the coalition, and can be easily understood by the members, which is important when charging end-users \cite{bonbright1988principles}. The D-NEM mechanism proposed in this paper augments the favorable features in \cite{Chakraborty&Poolla&Varaiya:19TSG} by including individual {\em surplus} and community {\em social welfare} as part of the design objectives of the community pricing in a decentralized optimization framework. When prosumer {\em surplus} is considered, standalone rational utility customers under NEM may achieve higher surpluses than the customers under \cite{Chakraborty&Poolla&Varaiya:19TSG}, which violates individual rationality, hence the cost-causation principle. 

To our best knowledge, D-NEM proposed here is the first energy community market mechanism that achieves efficiency, defined by social welfare maximization, and individual/group rationality under the cost-causation principle.

\subsection{Summary of Results and Contributions}
We propose {\em D-NEM}---an energy community market mechanism that sets the NEM price based on the aggregated DER within the community. D-NEM employs the utility's NEM import and export rates and applies them dynamically to individual members based on the community aggregated renewables. Unlike utility's NEM tariff whereby each prosumer faces differentiated import and export rates based on its own net consumption state, a prosumer under D-NEM faces only one rate depending solely on the community's aggregate net consumption state, an idea that was first articulated in \cite{Chakraborty&Poolla&Varaiya:19TSG}. 

The main differences between D-NEM and that in \cite{Chakraborty&Poolla&Varaiya:19TSG} are threefold. First, D-NEM prices are set {\em ex-ante} (rather than imposed {\em ex-post} in \cite{Chakraborty&Poolla&Varaiya:19TSG} and in most re-distribution schemes such as the {\em Shapley value}, {\em proportional rule}, and {\em nucleolus}) to induce community members' response that achieves the community's maximum welfare.  Second, D-NEM induces a community-level net-zero consumption zone whereby the community's aggregate flexible resources balance the shared renewables. Third, we incorporate a centralized battery energy storage system (BESS), and show how D-NEM market mechanism generalizes. 

We establish the following properties of D-NEM: 
\begin{enumerate}
\item Individual surplus maximizations lead to maximum community social welfare.
\item D-NEM and individual surplus maximization is a Nash equilibrium that ensures higher surplus for community members than the maximum surplus under DSO's NEM.
\item D-NEM satisfies the {\em cost-causation principle}.
\item The community is stable under D-NEM, \ie no subset of users would be better off if they jointly withdraw from the community and form their own.
\end{enumerate}

\par A key result in our work is that the maximum community welfare is decentrally achieved through a threshold-based pricing policy that is a function of the aggregated renewables.  Our empirical results use real residential data to construct an energy community, under which the welfare, price, and community operation under D-NEM are compared to communities under \cite{Chakraborty&Poolla&Varaiya:19TSG} and to standalone {\em optimal} DSO customers.

The extensions of this work over the preliminary version in \cite{Alahmed&Tong:23PESGM} are fourfold. First, a Stackelberg bi-level optimization is formulated to capture the interaction between the community operator and its members, and argue for the decentralized welfare maximization and {\em market equilibrium} (Theorem \ref{thm:Equilibrium}). Second, we show that D-NEM is a stabilizing allocation by proving that it satisfies {\em group rationality} (Theorem \ref{thm:GroupRationality}). Third, unlike the work in \cite{Alahmed&Tong:23PESGM}, which considered flexible demand and renewable distributed generation (DG) as available DERs, the work presented here incorporates a BESS (Sec.\ref{sec:DNEM2}), hence introducing more dynamics and inviting multiple time interval formulation. Through the market mechanism, the BESS is ultimately co-optimized with the other flexible community resources. Fourth, this work offers broader simulation results with a more nuanced analysis of the impact of co-optimizing the BESS with other DERs on the community's performance and price characteristics.

To argue that the energy community under D-NEM offers its members a surplus (benefit) higher than the maximum surplus when they autonomously face the DSO, we leverage our work in \cite{Alahmed&Tong:22IEEETSG,Alahmed&Tong&Zhao:23arXiv} on the analysis of optimal {\em standalone prosumer} response under the DSO's NEM tariff when the prosumer has renewable DG and flexible loads \cite{Alahmed&Tong:22IEEETSG} and when the prosumer adds a BESS \cite{Alahmed&Tong&Zhao:23arXiv}. Therefore, the maximum surplus under the optimal prosumer response in \cite{Alahmed&Tong:22IEEETSG,Alahmed&Tong&Zhao:23arXiv} is regarded as the {\em benchmark} surplus that the community operator attempts to compete with by proposing D-NEM.
\subsection{Notations and Nomenclature}

The use of notations here is standard. When necessary, boldface letters denote column vectors as in $\bm{x}=(x_1,\cdots, x_n)$ and $\bm{x}^\top$ its transpose. For vectors $\bm{x},\bm{y}$,  $\bm{x} \preceq \bm{y}$ is the element-wise inequality $x_i \le y_i$ for all $i$, and $[\bm{x}]^+, [\bm{x}]^-$ are the element-wise positive and negative parts of vector $\bm{x}$, \ie $[x_i]^+=\max\{0,x_i\}$, $[x_i]^- =-\min\{0,x_i\}$ for all $i$, and $\bm{x}= [\bm{x}]^+ - [\bm{x}]^-$. Lastly, we use $\mathbb{R}_+$ to denote the set of non-negative real numbers, \ie $\mathbb{R}_+:=\{ x\in \mathbb{R}: x\geq 0\}$. Key designations of symbols are given in Table \ref{tab:MajorSymbols}.

\begin{table}
\centering
\caption{Major designated symbols (alphabetically ordered).}
\label{tab:MajorSymbols}
\vspace{-0.2cm}
\resizebox{\columnwidth}{!}{%
\begin{tabular}{@{}ll@{}}
\toprule \midrule
Symbol                                                   & Description                                                            \\ \midrule
$b_i, b_{\Nc}$                                           & Storage output of member $i$ and community.             \\
$\underline{b}_{\Nc}, \overline{b}_{\Nc}$                & Central storage lower and upper limits.                                \\
$\chi$                                                   & Community operator's pricing policy.                                   \\
$\bm{d}_i, d_{\Nc}$                                      & Consumption of member $i$ and community.         \\
$\bm{d}_i^\psi$                                          & Consumption of member $i$ under consumption policy $\psi$.      \\
$\underline{\bm{d}}_i, \overline{\bm{d}}_i$              & Consumption bundle's lower \& upper limits of member $i$.             \\
$E$                                                      & Central storage size.                                                  \\
$\bm{f}_i(\cdot)$                                        & Inverse marginal utility function of customer $i$.                     \\
$g_i, g_{\Nc}$                                           & Generation of member $i$ and community.           \\
$\gamma$                                                 & Storage salvage value rate.                                            \\
$i,\Nc$                                                  & Index and set of community members.                                    \\
$k, \Kc$                                                 & Index and set of consumption devices.                                  \\
$\bm{L}_i(\cdot)$                                        & Marginal utility function of member $i$.                               \\
$P^\chi(\cdot), P^{\mbox{\tiny NEM}}(\cdot)$             & Payment function under policy $\chi$ and under NEM.                    \\
$\pi^+, \pi^-$                                           & NEM X buy (retail) and sell (export) rates.                            \\
$\psi$                                                   & Community members' consumption policy.                                 \\
$Q_{it}^{\chi}(\cdot), Q^{\mbox{\tiny NEM}}_{it}(\cdot)$ & Reward function of member $i$ under $\chi$ and NEM in time $t$. \\
$\rho, \tau$                                             & Storage discharging and charging efficiencies.                         \\
$S^\chi(\cdot), S^{\mbox{\tiny NEM}}(\cdot)$             & Surplus function under policy $\chi$ and under NEM.                    \\
$t,\Tc$                                                  & Index and set of time.                                                 \\
$U_i(\cdot)$                                             & Utility of consumption of member $i$.                                  \\
$W$                                                      & Community social welfare.                                              \\
$x_t$                                                    & Central storage state of charge at time $t$.                           \\
$\xi_i$                                                  & Share of community member $i$ in the central storage.                  \\
$z_i, z_{\Nc}$                                           & Net consumption of member $i$ and community. \\
$z_i^\psi$                                               & Net consumption of member $i$ under policy $\psi$.  \\ \midrule \bottomrule
\end{tabular}%
}
\end{table}

%% file: formulation_v2.tex
\begin{figure}
    \centering
    \includegraphics[scale=0.42]{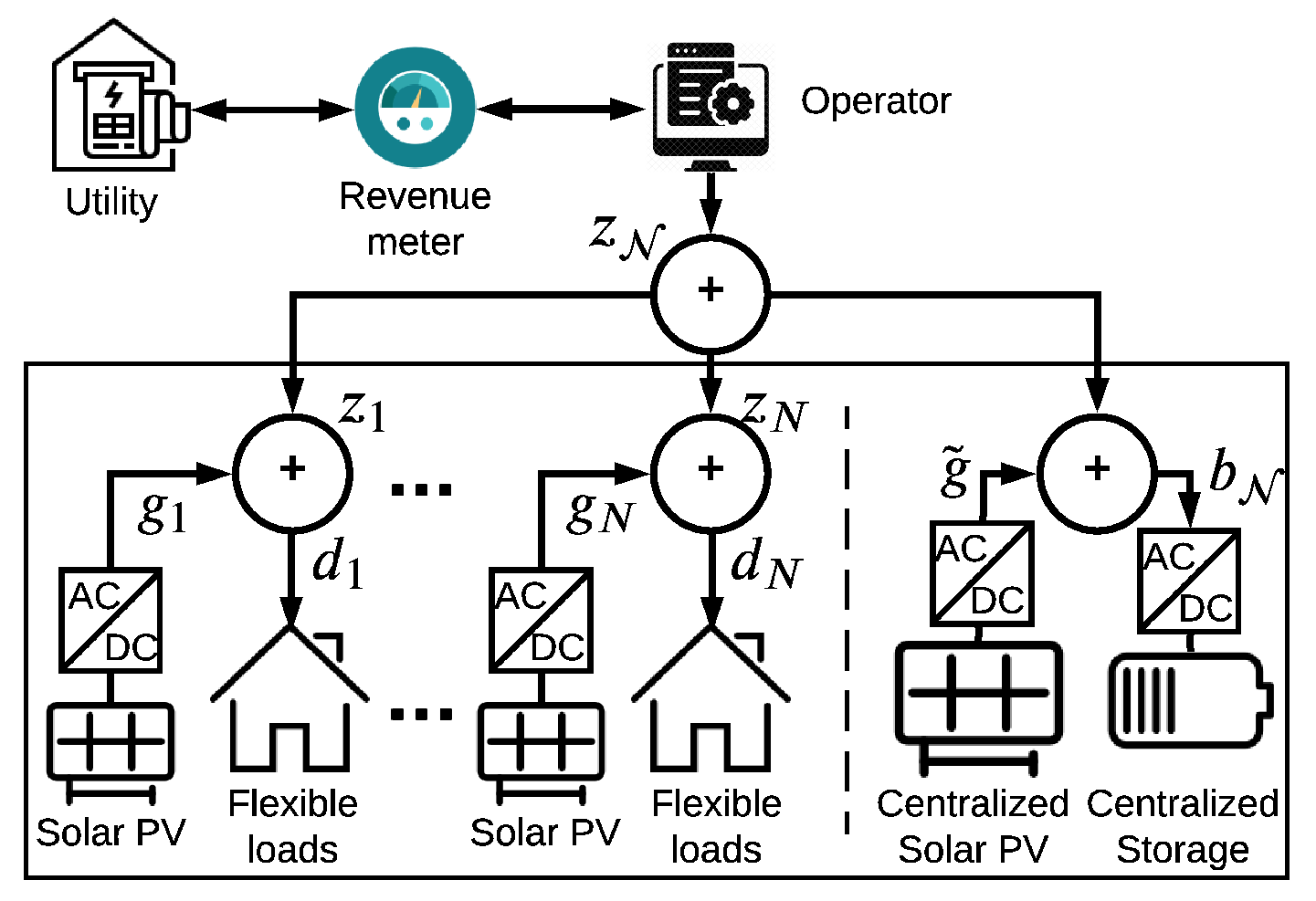}
    \vspace{-0.3cm}
    \caption{Energy community framework. The decentralized DERs include flexible consumption, renewables DG, and net consumption denoted by $d_i, g_i \in \mathbb{R}_+, z_i \in \mathbb{R}$, respectively. The centralized resources include solar PV $\tilde{g} \in \mathbb{R}_+$ and a BESS $b_{\Nc} \in \mathbb{R}$.}
    \label{fig:EnergyCommunity}
\end{figure}

We consider a community $\Nc$ of $N$ rational members,\footnote{The community members can be a composition of residential, commercial, industrial, etc.} each indexed by $i \in \Nc:=\{1,\ldots,N\}$, who pool and aggregate their resources behind a regulated utility revenue meter (Fig.\ref{fig:EnergyCommunity}) that implements an NEM policy design \cite{Alahmed&Tong:22IEEETSG}. The community resources include heterogeneous flexible demands, solar PVs, and a centrally scheduled BESS.\footnote{In Appendix B, we show how our analysis extends to communities with central renewables and even distributed BESS under mild assumptions.} The community members are subject to the market mechanism of a non-profit community operator, who performs energy and monetary transactions with the utility on behalf of the whole community (Fig.\ref{fig:EnergyCommunity}). Energy community members may choose to leave the community and be served autonomously by the DSO under its regulated NEM tariff.

The roles of a DSO and an energy community are different.  The DSO has the responsibility to operate the overall distribution system and serve its customers.  An energy community, on the other hand, only serves its customers financially. Through the profit-neutral energy community operator, community members pay for their net energy consumption and the community operator pays (gets compensated) for (by) the DSO for energy delivery to (from) its customers.  An energy community shares the underlying network with other customers, and its operator, unlike a microgrid operator, is not responsible for the physical network operation. Hence the network constraints are not part of the energy community's decision process. To ensure system reliability, the DSO may impose dynamic operating envelopes on the individual customers and the community \cite{Liu&Ochoa&Wong&Theunissen:22TSG}.  We do not consider such constraints here. See \cite{Alahmed&Cavraro&Bernstein&Tong:23AllertonArXiv} for further work involving operating envelopes.

To simplify the exposition and facilitate delivering the intuition of D-NEM, we first consider a community with flexible demands and solar PV. Then, in Sec.\ref{sec:DNEM2}, we add a BESS and show how D-NEM generalizes. 

\subsection{Member Resources}

Every member $i\in \Nc$ is assumed to have $K$ controllable devices operated by a home energy management system (HEMS), and indexed by $k \in \mathcal{K}=\{1,\ldots,K\}$. The {\em energy consumption bundle} for every member $i \in \Nc$ is defined by  $\bm{d}_i:=\left(d_{i1}, \ldots, d_{iK}\right)$, where
\begin{equation}\label{eq:Consumption}
    \bm{d}_i \in \mathcal{D}_i:=\{\bm{d}_i: \underline{\bm{d}}_i \preceq \bm{d}_i \preceq \overline{\bm{d}}_i\} \subseteq \mathbb{R}_{+}^K,
\end{equation}
and $\underline{\bm{d}}_i, \overline{\bm{d}}_i$ are the consumption bundle's lower and upper limits of member $i$, respectively. The {\em aggregate consumption} of the community is denoted by $d_{\Nc}:= \sum_{i\in \Nc} \bm{1}^\top \bm{d}_i$.

Members may have BTM {\em renewable DG} (Fig.\ref{fig:EnergyCommunity}), whose output is denoted by $g_i \in \mathbb{R}_{+}$ for every $i\in \Nc$. Community members, also, may own shares of the centralized (community) generation, which is already accounted for in $g_i$.\footnote{Owning shares from offsite renewable DG and considering the share's output as if it was generated BTM is enabled by VNEM programs \cite{NEMevolution:23NAS}.} The community's {\em aggregate DG} is $g_{\Nc}:= \sum_{i\in \Nc} g_i$.

The {\em net consumption} of every $i \in \Nc$ member $z_i \in \mathbb{R}$ and the community {\em aggregate net consumption} $z_{\Nc} \in \mathbb{R}$ (Fig.\ref{fig:EnergyCommunity}) are defined as
\begin{equation}\label{eq:NetCons}
    z_i:= \bm{1}^\top \bm{d}_i - g_i,~~~ z_{\Nc}:= \sum_{i =1}^N z_i = d_{\Nc}-g_{\Nc},
\end{equation}
where $z_i\geq 0$ and $z_i<0$ represent a {\em net-consuming} and {\em net-producing} member, respectively.

\subsection{Community Pricing and Member Surplus}

The energy community may have its own pricing rule different from that of the DSO for the benefit of the community and its members.  To this end, we denote the community's pricing policy by $\chi$ that defines the payment function 
$\bm{P}^{\chi}(\bm{z}):=(P^\chi_1(z_1),\ldots, P^\chi_N(z_N))$, where $P_i^\chi$ is the payment function for member $i$ under $\chi$, and $\bm{z}:=(z_1,\ldots,z_N)$.

For every $i \in \Nc$, the {\em surplus} of the community member is defined as
\begin{equation}\label{eq:Smember}
    S_i^\chi(\bm{d}_i,g_i):= U_i(\bm{d}_i)-P^\chi_i(z_i), ~~ z_i:= \bm{1}^\top \bm{d}_i - g_i,
\end{equation}
where $U_i(\bm{d}_i)$ is the utility function of consuming the bundle $\bm{d}_i$, which may have different functional forms for each $i\in \Nc$ based on their flexibility and desired comfort levels. For all $i\in \Nc, U_i(\bm{d}_i)$ is assumed to be concave, non-decreasing, continuously differentiable, and additive, therefore, $U_i\left(\bm{d}_i\right):=\sum_{k\in \mathcal{K}} U_{i k}\left(d_{ik}\right), \forall i \in \Nc$.
We denote the marginal utility function by $\bm{L}_i:=\nabla U_i$, and its inverse, for every $i \in \Nc$ and $k \in \mathcal{K}$, by $f_{ik}:= L^{-1}_{ik}$.

\subsection{Standalone Prosumer Under DSO Pricing}
Each community member may compare its benefits as a member of the community with the benchmark if it is a standalone customer outside the community under the DSO's regulated NEM tariff. In particular, the generic NEM X tariff \cite{Alahmed&Tong:22IEEETSG} is defined by NEM payment function $P^{\mbox{\tiny NEM}}$ given by 
\begin{equation}\label{eq:Pbenchmark}
P^{\mbox{\tiny NEM}}(z_i) = \pi^+[z_i]^+-\pi^-[z_i]^-,
\end{equation}
where $\pi^+ \in \mathbb{R}_+$ and $\pi^-\in \mathbb{R}_+$ are the {\em buy (retail) rate}, and {\em sell (export) rate}, respectively, with $\pi^- \leq \pi^+$.\footnote{Setting $\pi^- \leq \pi^+$ is in line with current and evolving NEM practices \cite{Alahmed&Tong:22EIRACM,NEMevolution:23NAS}. It also prevents risk-free price arbitrage. We ignore the fixed charge in the NEM X payment, which is used by the DSO to recover network and connection costs, as it can be directly passed by the operator to its members to match their benchmark fixed charge.} In practice, (\ref{eq:Pbenchmark}) is referred to as {\em NEM 1.0} if $\pi^- = \pi^+$, and {\em NEM 2.0} and beyond if $\pi^-<\pi^+$ \cite{Alahmed&Tong:22EIRACM}. $\pi^+$ and $\pi^-$ can be temporally fixed (flat rates) or varying (ToU rates). With little loss of generality, we assume that members' decisions have the same timescale as that of the NEM {\em netting frequency} \cite{Alahmed&Tong:22EIRACM}.
  
If member $i$ is a standalone prosumer of the DSO, its surplus from consumption $\bm{d}_i$ and net consumption $z_i := {\bf 1}^\top \bm{d}_i-g_i$ is
\begin{equation}\label{eq:Sbench}
S_i^{\mbox{\tiny NEM}}(\bm{d}_i,g_i)=U_i(\bm{d}_i)-P^{\mbox{\tiny NEM}}(z_i).
\end{equation}

\subsection{Axiomatic Community Pricing}
We specialize the general axiomatic cost-allocation principle to the community pricing rule, requiring that community pricing satisfies the following four axioms.

\begin{axiom}[Uniform payment function]\label{ax:equity}
    The market mechanism has a uniform payment function if, for any two community members $i,j \in \Nc, i\neq j$, having $z_i =z_j$ results in $P^{\chi}_i=P^{\chi}_j$.
\end{axiom}
For the community pricing problem considered here, the uniform payment function is equivalent to the more general {\em equal treatment of equals} principle of the coalition cost allocation axiom.

We present the next few axioms based on the uniform payment function $P^\chi$.

\begin{axiom}[Monotonicity and cost-causation]\label{ax:monotonicity}
   The payment function $P^\chi$ is monotonic and $P^\chi(0)=0$.
\end{axiom}
The monotonicity in axiom \ref{ax:monotonicity} ensures that when any two members $i,j \in \Nc, i\neq j$ have the same net consumption sign $z_i z_j \geq 0$  and $|z_i| \geq |z_j|$, then the member with higher net consumption (production) should have higher payment (compensation), \ie $|P^{\chi}(z_i)|\geq |P^{\chi}(z_j)|$. Also axiom \ref{ax:monotonicity} ensures cost-causation because if a member $i\in \Nc$ causes cost, \ie $z_i>0$, then it is penalized for it, hence $P^\chi(z_i)\geq 0$, whereas if it mitigates cost, \ie $z_i<0$, then it is rewarded for it, hence $P^\chi(z_i)\leq 0$.

\begin{axiom}[Individual rationality]\label{ax:rationality}
   The surplus of every $i\in \Nc$ member should be no less than its benchmark as a standalone customer under DSO's NEM tariff, \ie $S_i^\chi \ge S_i^{\mbox{\tiny NEM}}$.
\end{axiom}
The individual rationality axiom requires the pricing policy $\chi$ to make joining the community advantageous for every member by offering them a surplus that is competitive to the surplus achieved under the DSO's NEM tariff.

\begin{axiom}[Profit neutrality]\label{ax:ProfitNeutrality}
The community pricing rule must ensure the operator's profit neutrality, \ie  
\[\sum_{i\in \Nc} P^\chi_i(z_i) = P^{\mbox{\tiny NEM}}(\sum_{i\in \Nc} z_i).\]
\end{axiom}
Axiom \ref{ax:ProfitNeutrality} is consistent with the energy community definition under \cite{CitizenEC:EuroCommWebsite} requiring the operator to be budget-balanced.

\begin{definition}[Cost-causation principle]\label{def:CausationPrinciple}
    A community payment function $\bm{P}^\chi$ conforms with the {\em cost causation principle} if $\bm{P}^\chi \in \Ac:=\{\bm{P}: \text{Axioms \ref{ax:equity}--\ref{ax:ProfitNeutrality} are satisfied}\}$.
\end{definition}

Note that using the {\em surplus} rather than {\em payment} (cost) to define {\em individual rationality} is more general as the surplus function includes both the consumption preferences of the prosumer and its cost (payment). Interestingly, the DSO's NEM tariff does not conform with the cost causation principle $P^{\mbox{\tiny NEM}} \notin \Ac$, as $P^{\mbox{\tiny NEM}}$ does not satisfy profit neutrality (Axiom \ref{ax:ProfitNeutrality}). In fact, most of the {\em ex-post} allocation mechanisms such as Shapley value, proportional rule, and the nucleolus of a coalition game do not conform with the cost causation principle, as they, in particular, do not satisfy Axioms \ref{ax:equity}--\ref{ax:monotonicity}.   

%% file: DECWel_v2.tex
This section introduces a decentralized welfare maximization framework through a Stackelberg bi-level optimization, where the community operator is the {\em leader} who optimizes its pricing rule $\chi$ to maximize community welfare and the community members are {\em followers}, maximizing individual consumption benefits given the operator's pricing rule.

A clarification of notations is helpful.  Generically, the operator's pricing policy is denoted by $\chi$ and members consumption policy by $\psi$. We use $\chi_\psi$ for the policy of the operator given member's consumption policy $\psi$ and $\psi_\chi$ the consumption policy given $\chi$.  We use superscript $\sharp$ for {\em conditionally optimal policy}, \ie $\chi^\sharp_\psi$ is the optimal pricing policy given $\psi$, and likewise $\psi^\sharp_\chi$ the optimal consumption policy given $\chi$.  

We begin with the formulation of the lower-level optimization of community members followed by the upper-level optimization of the operator.

\subsection{Members' Optimal Consumption Policy $\psi_\chi^\sharp$} 
With uniform pricing, all members face the same payment rule defined by $P^\chi$.  Here we drop the member index and consider the consumption policy of a generic community member $\psi_\chi$ given the operator's pricing rule $\chi$.  To this end, we further assume that all community members are price-taking surplus maximizers with optimal consumption policy $\psi_\chi^\sharp$ (given $\chi$) is defined by $\psi^\sharp_\chi: g \stackrel{P^\chi}{\mapsto} (\bm{d}^\psi,z^\psi)$, where
\begin{align}\label{eq:LowerLevel}
 (\bm{d}^\psi,z^\psi) :=\underset{\bm{d}\in \mathbb{R}_+^K, z\in \mathbb{R}}{\operatorname{argmax}}&~~~ S^\chi(\bm{d},g):=U(\bm{d})-P^\chi(z)\nn\\
 		~~~~\text{subject to}&~~~ z= \bm{1}^\top \bm{d} - g\\
		&~~~	\bm{d} \in \mathcal{D}:=\{\bm{d}: \underline{\bm{d}} \preceq \bm{d} \preceq \overline{\bm{d}}\}.\nn
\end{align}
 
Note that the community member schedules its consumption based on its DER $g$ that includes both member's private and shared resources and is modeled as stochastic.  Note also that, although the above formulation is presented as a single interval optimization, it can be generalized naturally to multi-interval optimization when battery resources are involved.

\subsection{Operator's Optimal Pricing Policy $\chi_\psi^\sharp$} 

Given an individual member's consumption policy $\psi$, the community operator optimizes the pricing rule (policy) $\chi$ that defines the community payment function $P^\chi$.  The timescale for the operator's decision process may not match with that of members' consumption decisions.  The community pricing rule may be set on a daily, weekly, or monthly basis whereas the consumption decisions are made at the minute, hourly, or daily basis.   Because of such mismatch, the pricing decision of the operator is fundamentally stochastic without knowing the realizations of renewable.

We assume that the operator's pricing objective is maximizing the expected community social welfare defined by the total surplus of its members

\begin{equation*}
W^{\chi,\psi}:=\sum_{i\in \Nc} \mbbE[S_i^{\chi}(\psi(g_i),g_i)],
\end{equation*}
where the expectation is taken over member DG ($g_i$).  Under Axiom \ref{ax:ProfitNeutrality} (profit neutrality), the (conditional) welfare maximizing pricing policy $\chi^\sharp_\psi$  (given $\psi$) is defined by $\chi^\sharp_\psi: \bm{g}:=(g_1,\ldots, g_N) \mapsto P^{\chi}$ where

\begin{align}\label{eq:UpperLevel}
 P^{\chi} := \underset{P(\cdot) \in \Ac}{\operatorname{argmax}}& \Bigg(W^{\chi_\psi} = \mbbE\Big[\sum_{i\in \Nc} U_i(\bm{d}_i^\psi) - P^{\mbox{\tiny NEM}}(z_{\Nc})\Big]\Bigg)\nn\\
 \text{subject to}&~~~ z_{\Nc} = \sum_{i\in \Nc} \bm{1}^\top \bm{d}_i^\psi- g_i,
\end{align}

where $\bm{d}_i^\psi$ is the consumption generated by the consumption policy of member $i$.  Unfortunately, the optimization above is infinite-dimensional; it cannot be solved easily without exploiting the NEM structure of the regulated rate. 

\subsection{Stackelberg Equilibrium}
The above bi-level optimization defines the so-called {\em Stackelberg equilibrium} (or Stackelberg strategy).  Specifically, ($\chi^\ast,\psi^\ast$) is a Stackelberg equilibrium if (a)  all $\chi \in \Xmsc,   S^\chi(\psi^\ast(g),g) \ge S^\chi(\psi(g),g)$ for all $\psi \in \Psi$; (b) for all $\psi \in \Psi, W^{\chi^\ast,\psi^\ast} \ge \mbbE[\sum_iS_i^\chi(\psi^\ast(g_i),g_i)]$.  In other words, the Stackelberg equilibrium is the optimal community pricing when community members optimally respond to the community pricing.

%% file: DNEM1_v2.tex
Here, we propose D-NEM---a cost-causation-based market mechanism that induces community members to achieve the maximum community welfare, and equilibrium to the bi-level optimization (\ref{eq:LowerLevel})-(\ref{eq:UpperLevel}).

\subsection{Dynamic NEM: Optimal Community Pricing}\label{subsec:DNEM}
Given the aggregate renewable DG $g_\Nc$, the operator announces the community pricing rule according to D-NEM policy shown below.

\begin{policy*}\label{alloc:ProfitNeutral}
Under D-NEM, the energy community pricing policy $\chi^\ast: \gbf \mapsto P^{\chi^\ast}(z)=\pi^\ast(g_\Nc)\cdot z$  is threshold-based with
\begin{align}
\pi^\ast(g_\Nc)&=\begin{cases}   \pi^+&, g_\Nc<f_\Nc(\pi^+) \\ \pi^z(g_\Nc) &, g_\Nc\in[f_\Nc(\pi^+),f_\Nc(\pi^-)] \\ \pi^-&,
 g_\Nc>f_\Nc(\pi^-) \end{cases}\label{eq:PricingMechanism}
 \end{align}
 where the thresholds $f_\Nc(\pi^+)\leq f_\Nc(\pi^-)$ are given by
 \begin{align}\label{eq:thresholds}
 f_\Nc(y):=\sum_{i \in \Nc} \sum_{k \in \Kc} \max\{\underline{d}_{ik}, \min\{f_{ik}(y),\bar{d}_{ik}\}\}
\end{align}
and the price $\pi^z(g_\Nc):=\mu^\ast(g_\Nc)$ is the solution of:
 \begin{equation}\label{eq:NetZeroPrice}
 f_\Nc(\mu) = g_\Nc.
\end{equation}
\end{policy*}
A numerical example is provided in Appendix A to show how the D-NEM price is computed.
\subsubsection{Structural Properties of D-NEM}
\begin{figure}
    \centering
    \includegraphics[scale=0.40]{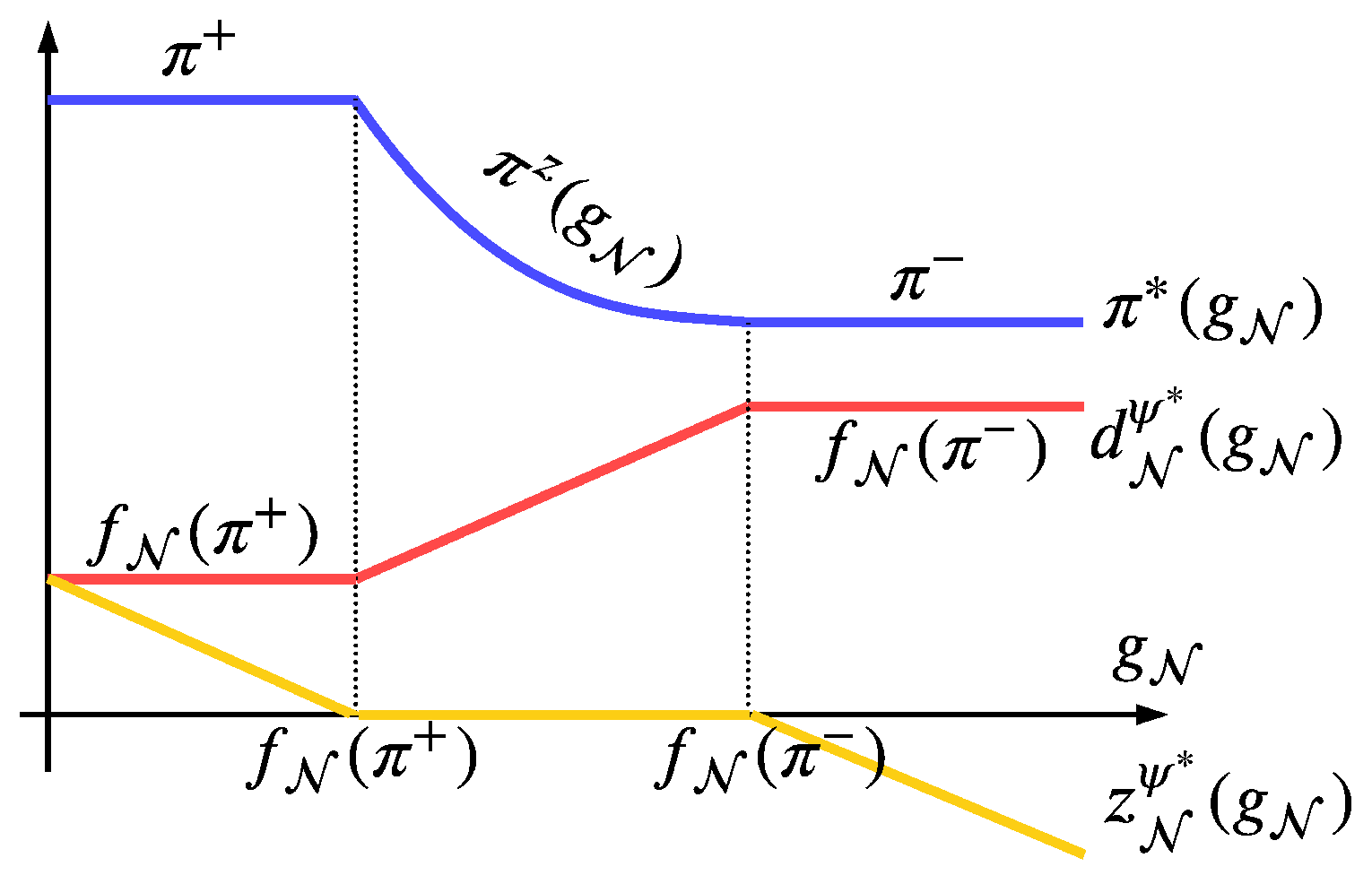}
    \vspace{-0.3cm}
    \caption{Depiction of community prices and optimal operation under D-NEM.}
    \label{fig:DNEM_R1}
\end{figure}

The dynamic pricing policy has an appealing threshold-based resource-aware structure, that announces community prices based on the aggregate renewables level compared to two renewable-independent thresholds $f_\Nc(\pi^+)$ and $f_\Nc(\pi^-)$, which can be computed {\em apriori}. Remarkably, D-NEM only requires the aggregate renewables $g_\Nc$ rather than the full $\bm{g}$, which implies that the operator does not need to forecast/measure the private BTM renewables. Fig.\ref{fig:DNEM_R1} shows the community price and operation under D-NEM after accounting for members' response  $\psi^\ast_i:g_i \stackrel{P^{\chi^\ast}_i(z_i)}{\mapsto} (\bm{d}^{\psi^\ast}_i,z_i^{\psi^\ast}), \forall i\in \Nc$, as will be shown in Sec.\ref{subsec:OptConsPol}. In response to D-NEM, the community aggregate consumption and net consumption are defined as $d^{\psi^\ast}_\Nc:= \sum_{i\in \Nc} \bm{1}^\top \bm{d}^{\psi^\ast}_i$ and $z^{\psi^\ast}_\Nc:= d^{\psi^\ast}_\Nc - g_\Nc$, respectively.

As shown in Fig.\ref{fig:DNEM_R1}, D-NEM passes the DSO's $\pi^+$ and $\pi^-$ when $g_\Nc < f_\Nc(\pi^+)$ and $g_\Nc > f_\Nc(\pi^-)$, respectively. When ($g_\Nc \in [f_\Nc(\pi^+),f_\Nc(\pi^-)]$), D-NEM dynamically charges members by the Lagrangian multiplier satisfying the Karush-Kuhn-Tucker (KKT) condition of $d^{\psi^\ast}_\Nc(g_\Nc)=g_\Nc$. 

\begin{proposition}[Net-zero zone price $\pi^z(g_\Nc)$]\label{corol:NZprice}
When $g_\Nc \in [f_\Nc(\pi^+),f_\Nc(\pi^-)]$, $\pi^\ast(g_\Nc) =\pi^z(g_\Nc) \in [\pi^-, \pi^+]$ and the price $\pi^z(g_\Nc)$ monotonically decreases with increasing $g_\Nc$. 
\end{proposition}
\noindent {\em Proof:} See Appendix C.\hfill$\blacksquare$\\
Proposition \ref{corol:NZprice} shows that the net-zero zone price $\pi^z(g_\Nc)$ is bounded by the NEM X rates, and dynamically decreases with increasing $g_\Nc$ to incentivize demand increases and keep the community off the grid. 

It is worth noting that, unlike market mechanisms that clear the community price based on exogenously determining the {\em roles} of community prosumers as {\em buyers} and {\em sellers} \cite{Sorin&Bobo&Pinson:19TPS,Hu&Zhang:21TSG}, D-NEM liberates prosumers to choose their {\em role}.

\subsubsection{Intuitions of D-NEM}
The resource-aware D-NEM pricing policy is intuitive as it responds to the increasing community generation by dynamically reducing the price. In contrast, the regulated NEM X tariff has predetermined prices independent of the available renewables. 

To a community member, D-NEM offers a significant advantage over the DSO's regulated NEM. When a member $i$ is net-producing $z_i<0$, it is compensated at a price no less than the NEM sell rate $\pi^\ast(g_\Nc)\geq \pi^-$ it would face outside the community. Also, when a member $i$ is net-consuming $z_i\geq 0$, it is charged at a price no more than the NEM buy rate it would face outside the community $\pi^\ast(g_\Nc)\leq \pi^+$. Thus, the market under D-NEM is strictly advantageous to the net-producer ({\em seller}) if $\pi^\ast(g_\Nc)>\pi^-$ and to the net-consumer ({\em buyer}) if $\pi^\ast(g_\Nc)<\pi^+$. This intuition supports but does not fully imply, the individual rationality axiom.

\subsubsection{Privacy Considerations}
To announce the D-NEM price, the operator needs to compute the two thresholds ($f_\Nc(\pi^+),f_\Nc(\pi^-)$) in (\ref{eq:PricingMechanism}) and compare them to the aggregate generation in the community $g_\Nc$. Both thresholds can be computed without compromising members' privacy. Indeed, given the exogenous and known DSO's NEM prices ($\pi=(\pi^+,\pi^-)$), each community member $i\in \Nc$ submits a value for every device $k$ based on $\max\{\underline{d}_{ik}, \min\{f_{ik}(\pi),\bar{d}_{ik}\}\}$, which allows the operator to solve (\ref{eq:thresholds}). A subtle case arises when the community is in the net-zero zone ($g_\Nc\in[f_\Nc(\pi^+),f_\Nc(\pi^-)]$), and the operator needs to compute $\pi^z(g_\Nc)$ to which we address by propagating samples of the member locally-computed values $\max\{\underline{d}_{ik}, \min\{f_{ik}(x),\bar{d}_{ik}\}\}, x\in [\pi^-,\pi^+]$ to the operator. Such sampling is not hard given that the net-zero zone price is bounded by the exogenous DSO rates (Proposition \ref{corol:NZprice}).

\subsection{Members' Optimal Consumption Policy}\label{subsec:OptConsPol}
Given the announced D-NEM price and payment rules, every $i \in \Nc$ member maximizes its surplus by optimally scheduling its consumption (lower level problem in (\ref{eq:LowerLevel})) as
\begin{align} \label{eq:LowerLevelOptimal}
(\bm{d}^{\psi^\ast}_i,z_i^{\psi^\ast})\hspace{-0.05cm}:=~&\hspace{-0.43cm} \underset{\bm{d}_i \in \mathbb{R}_{+}^K,z_i \in \mathbb{R}^K}{\operatorname{argmax}}  \hspace{-0.05cm}\bigg(S_i^{\chi^\ast}(\bm{d}_i,z_i) := U_i\left(\bm{d}_i\right)- P^{\chi^\ast}_i(z_i)\bigg) \nn \\
& \text {subject to } \quad  z_i = \mathbf{1}^{\top} \bm{d}_i-g_i\\
&\hspace{2cm} \underline{\bm{d}}_i \preceq \bm{d}_i \preceq \overline{\bm{d}}_i \nn.
\end{align} 
A straightforward, yet crucial, implication of D-NEM compared to the DSO's NEM, is that the member's optimal consumption becomes solely a function of D-NEM price, as
\begin{equation}\label{eq:optconsi}
    d^{\psi^\ast}_{ik}(\pi^\ast) = \max \{ \underline{d}_{ik},\min\{f_{ik}(\pi^\ast),\bar{d}_{ik}\}\}, \forall k \in \Kc, \forall i\in \Nc,
\end{equation}
which can be abstractly written in the vector format $\bm{d}^{\psi^\ast}_{i}(\pi^\ast):=(d^{\psi^\ast}_{i1},\ldots,d^{\psi^\ast}_{iK})$. The solution of (\ref{eq:LowerLevelOptimal}) results in the community member's maximum surplus $S_i^{\chi^\ast}(\bm{d}^{\psi^\ast}_{i},z^{\psi^\ast}_{i})$, where $z^{\psi^\ast}_{i}(\pi^\ast):= \bm{1}^\top \bm{d}^{\psi^\ast}_{i}(\pi^\ast) - g_i$. 

\subsection{Stackelberg Equilibrium and Properties}\label{subsec:Stackelberg}
It turns out that the D-NEM pricing rule achieves an equilibrium for the operator and community member interaction.
\begin{theorem}[Equilibrium of the bi-level optimization and welfare optimality]\label{thm:Equilibrium}
The solution ($\chi^\ast,\psi^\ast$) is an equilibrium to the bi-level optimization (\ref{eq:LowerLevel})-\ref{eq:UpperLevel}). Also, the community maximum social welfare under centralized operation is achieved by the equilibrium solution, \ie
    \begin{align} \label{eq:SurplusMax_model2}
(\bm{d}_1^{\psi^\ast},\ldots,\bm{d}_N^{\psi^\ast},\bm{z}^{\psi^\ast})\hspace{-0.05cm} = \hspace{-0.18cm}\underset{(\bm{d}_i,\ldots,\bm{d}_N),\bm{z}}{\rm argmax}& \hspace{-0.0cm} \mathbb{E}\big[\sum_{i\in \Nc} U_i(\bm{d}_i)-P^{\mbox{\tiny NEM}}(z_\Nc)\big]\nn \\ \text{subject to} &~~ z_\Nc = \sum_{i\in \Nc} \bm{1}^\top \bm{d}_{i}-g_i \nn\\&\hspace{0.4cm}\underline{\bm{d}}_i\preceq \bm{d}_i \preceq \overline{\bm{d}}_i,~ \forall i\in \Nc,\nn
\end{align}
\end{theorem}
\noindent {\em Proof:} See Appendix C.\hfill$\blacksquare$\\

The equilibrium in Theorem \ref{thm:Equilibrium} is proved by solving an upper bound of the community operator optimization that does not constrain the payment to the set of cost-causation-based rules.
A major result of Theorem \ref{thm:Equilibrium} is that the maximum welfare is decentrally achieved and that D-NEM satisfies the axioms of the cost-causation principle (Definition \ref{def:CausationPrinciple}). Therefore, in addition to inducing community members to achieve social welfare optimality, D-NEM grants them surplus levels that are at least equal to their maximum surpluses under the DSO's NEM X tariff derived in \cite{Alahmed&Tong:22IEEETSG} (individual rationality). Interestingly, this also applies to community members who do not own renewable DG.

It is worth noting that many re-distribution schemes do not satisfy the cost-causation principle, including the equal surplus division \cite{Driessen&Funaki:1991}, proportional rule \cite{Chakraborty&Baeyens&Khargonekar:18TPS}, and Shapley value \cite{Chakraborty&Poolla&Varaiya:19TSG}. Additionally, a major difference between D-NEM and the other notable re-distribution schemes is that D-NEM prices are set \textit{ex-ante} to induce community members to achieve a favorable objective (Theorem \ref{thm:Equilibrium}), whereas the re-distribution schemes impose payments/allocations to community members \textit{ex-post}.

\subsection{Stability under D-NEM}
Group rationality is an extension of individual rationality (Axiom \ref{ax:rationality}), under which no subset of users would be better off if they jointly withdrew and formed a new community. Group rationality assures the {\em stability} of the community, as no sub-communities can form \cite{Abada&Ehrenmann&Lambin:20EP}. 

\begin{theorem}[Group rationality under D-NEM]\label{thm:GroupRationality}
 Group rationality is satisfied under D-NEM, \ie for $\Hc \subseteq \Sc \subseteq \Nc$, denote the surplus of a member who join communities $\Sc$ and $\Hc$ by $S_{i,\Sc}^{\chi^\ast}$ and $S_{i,\Hc}^{\chi^\ast}$, respectively, then under D-NEM we have
 \begin{equation}\label{eq:GrpRat}
     \sum_{i \in \Hc} S_{i,\Sc}^{\chi^\ast}(\bm{d}^{\psi^\ast}_i,z_i^{\psi^\ast}) \geq \sum_{i \in \Hc} S_{i,\Hc}^{\chi^\ast}(\bm{d}^{\psi^\ast}_i,z_i^{\psi^\ast}).
 \end{equation}
\end{theorem}
\noindent {\em Proof:} See Appendix C.\hfill$\blacksquare$\\
In case $\sum_{i \in \Hc} S_{i,\Sc}^{\chi^\ast}<\sum_{i \in \Hc} S_{i,\Hc}^{\chi^\ast}$, community $\Hc$ is better off withdrawing from its mother community $\Sc$. In coalitional game theory, when group rationality is achieved (Theorem \ref{thm:GroupRationality}), in addition to individual rationality (Theorem \ref{thm:Equilibrium}) and efficiency (full re-distribution of coalition value), D-NEM is a {\em stabilizing allocation}.

%% file: DNEM2_v2.tex
In the previous section, we proposed D-NEM for decentralized welfare maximization when the community has distributed renewable PV and flexible demands. Here, we add a centrally scheduled BESS\footnote{With mild assumptions, our results apply to distributed BESS. See Appendix B.} co-optimized with flexible demands via D-NEM to maximize the community's welfare. In the sequel, we reformulate the problem to a multi-interval one and introduce the BESS-related parameters and variables, then, we introduce the generalized D-NEM mechanism.

\subsection{Energy Community with BESS}
\subsubsection{BESS System Modeling}
A BESS system with capacity $E>0$ is added to the community to provide higher flexibility and more self-consumption of renewables. To account for the BESS and the related dynamics, we consider a multi-time interval formulation under which the community's flexible resources are sequentially scheduled over a finite horizon $T$ indexed by $t\in \Tc:=\{0,\ldots,T\}$. At each stage $t\in \Tc$, let $x_t \in [0,E]$ denote the BESS state of charge (SoC) at the beginning of stage $t$. For every $t=0,\ldots,T-1$, $b_{\Nc t}:=\left[b_{\Nc t}\right]^{+}-\left[b_{\Nc t}\right]^{-} \in [-\underline{b}_\Nc ,\overline{b}_\Nc ]$ denote the storage output with $\underline{b}_\Nc $ and $\overline{b}_\Nc $ as the maximum energy discharging and charging rates, respectively. The battery is charged when $b_{\Nc t}>0$ and discharged when $b_{\Nc t}<0$. The charging and discharging efficiencies are denoted by $\tau \in (0,1]$ and $\rho \in (0,1]$, respectively. Given the storage level and output at stage $t$, the SoC evolves as per the following
\begin{equation}\label{eq:SoCEvolution}
    x_{t+1}=x_{t}+\tau\left[b_{\Nc t}\right]^{+}-\left[b_{\Nc t}\right]^{-} / \rho, \quad t=0, \ldots, T-1,
\end{equation}
with initial SoC $x_0=0$. 

A community member $i \in \Nc $ may own a share $\xi_i\in \Xi=\{\xi_i\in [0,1]:\sum_{i\in \Nc }\xi_i=1\}$ of the BESS capacity $e_i := \xi_i E$, which makes their output,\footnote{Note that $b_{it}$ is virtually accounted to the member as if it was a BTM BESS, which is the practice in VNEM \cite{NEMevolution:23NAS}.} for $t=0, \ldots, T-1$, $b_{it}=\xi_i b_{\Nc t} \in [-\underline{b}_i,\overline{b}_i] \in \mathbb{R}$, where $\underline{b}_i:=\xi_i\underline{b}_\Nc $ and $\overline{b}_i:=\xi_i \overline{b}_\Nc $ are the maximum energy discharging and charging
rates of member $i$'s storage share, respectively.\footnote{The model allows for adjusting the storage shares $\xi_i \in \Xi$ every $T$ period.}
\par For every $t=0,\ldots,T-1$, the {\em net consumption} of every member $i \in \Nc $ and the community's {\em aggregate net consumption} are, respectively, redefined as 
\begin{equation*}\label{eq:NetConsWithES}
    z_{it}:= \bm{1}^\top \bm{d}_{it}+b_{it}-g_{it},~~z_{\Nc t}:=\sum_{i\in \Nc } z_{it}:= d_{\Nc t} + b_{\Nc t} - g_{\Nc t}.
\end{equation*}

\subsubsection{Payment, Surplus, and Reward}
Given $z_{it}$ and $z_{\Nc t}$, we use the same payment and surplus functions introduced before with added time dimension. The surplus function $S_{it}^\chi$ does not take into account storage gains/losses due to charging/discharging. Therefore, we generalize it to the {\em reward function} $Q_{it}^\chi$ of consuming $\bm{d}_i$ and net-consuming $z_i:=\bm{1}^\top \bm{d}_i - g_i$ as, for every $i\in \Nc $ and for $t \in[0, T-1]$,
\begin{equation}\label{eq:MemberReward}
    Q_{it}^{\chi}(\bm{d}_{it},g_{it}):= S^{\chi}_{it}(\bm{d}_{it},g_{it}) + \gamma (\tau [b_{it}]^+-[b_{it}]^-/\rho ),
\end{equation}
where the second term on the right-hand side is the reward (cost) of charging (discharging) the storage share, both incurred at the salvage value rate $\gamma \in \mathbb{R}_+$. We assume $\gamma \in [\frac{1}{\tau} \max\{(\pi_t^-)\}, \rho \min\{(\pi_t^+)\}]$ to avoid trivial storage scheduling \cite{Alahmed&Tong&Zhao:23arXiv}. Similar to (\ref{eq:MemberReward}), the {\em reward} of a benchmark prosumer (standalone) is, for every $i\in \Nc $ and for $t \in[0, T-1]$,
\begin{equation*}\label{eq:BenchReward}
    Q_{it}^{\mbox{\tiny NEM}}(\bm{d}_{it},b_{it},g_{it}):= S^{\mbox{\tiny NEM}}_{it}(\bm{d}_{it},g_{it}) + \gamma (\tau [b_{it}]^+-[b_{it}]^-/\rho ).
\end{equation*}
Note that the customer faces $\gamma$ with and without the community, which builds on the assumption that both the operator and the benchmark prosumer place the same value on their stored energy given that they face the same NEM X tariff.

The operator under BESS designs community pricing with the objective of maximizing the expected {\em community social welfare}; defined here as the aggregate reward of its members over the $T$ intervals:
\begin{equation}\label{eq:CommReward}
W^{\chi,\psi}:= \sum_{t=0}^{T-1} \sum_{i\in \Nc} \mbbE [Q_{it}^{\chi}(\psi(g_{it}),g_{it})].
\end{equation}

\subsubsection{Resource Co-Optimization}\label{subsec:DERcooptimization}
If the community welfare was to be centrally maximized, the operator would
have formulated the storage-consumption co-optimization as a $T$-stage Markov decision process (MDP).  The state $s_t :=(x_t, g_{\Nc t}) \in \mathcal{S}$ of the MDP in interval $t$ includes the battery SoC $x_t$ and aggregate renewables $g_{\Nc t}$, whose evolution is defined by (\ref{eq:SoCEvolution}) and the exogenous Markov random process $(g_{\Nc t})$. The initial state is denoted by $s_0=(x_0, g_{\Nc 0})$. An MDP {\em policy} $\mu := (\mu_0,\ldots,\mu_{T-1})$ is a sequence of decision rules, $s_t \stackrel{\mu_t}{\rightarrow} u_t := (\{\bm{d}_{it}\}_{i=1}^N,b_{\Nc t},\{z_{it}\}_{i=1}^N)$, that specifies consumption and storage operation in each $t$.

\par  Therefore, if the community resources were centrally operated, the operator would have solved the following:\vspace{-0.1cm}
\begin{subequations} \label{eq:RewardMax_model2}
\begin{align} 
\underset{\mu = (\mu_0,\ldots,\mu_{T-1})}{\text { maximize}} &~~ W^{\mbox{\tiny NEM}}(s_t,u_t):=\mbbE_{\mu}\Big[\sum_{t=0}^{T-1} \sum_{i\in \Nc} Q_{it}^{\chi}(s_t,u_t)\Big] \\\text { subject to~~} & \text{for all}~ t=0,\ldots, T-1, \\&   (\ref{eq:SoCEvolution}) \\&    g_{\Nc (t+1)} \sim F_{g_{\Nc (t+1)}|g_{\Nc t}} \\&    0 \leq x_{t} \leq E   \label{eq:SoCcapacity}\\&    -\underline{b}_\Nc  \leq \left[b_{\Nc t}\right]^{+}-\left[b_{\Nc t}\right]^{-} \leq \overline{b}_\Nc    \\&    \bm{\underline{d}}_i \preceq \mathbf{d}_{it} \preceq \bm{\overline{d}}_i,~~ i=1,\ldots, N\\&   
P_{t}^{\mbox{\tiny NEM}}(\sum_{i\in \Nc } z_{i t}) = \sum_{i\in \Nc } P^{\chi}_{it}(z_{it}) \\&
 s_{0}=(x_0,g_{\Nc 0}),
\end{align} 
\end{subequations}
where $F_{g_{\Nc (t+1)}|g_{\Nc t}}$ is the conditional distribution of $g_{\Nc (t+1)}$ given $g_{\Nc t}$, and the expectation is taken over the exogenous stochastic aggregate generation $(g_{\Nc t})$. Note that, by definition, $[b_{\Nc t}]^+ \cdot [b_{\Nc t}]^-=0$.

The continuous-state space MDP defined above is intractable to solve. A practical solution is to implement a Model Predictive Control (MPC) where one solves a sequence of one-shot (static) sliding window optimizations \cite{Chen&Wang&Heo&Kishore:13TSG}.  Such a strategy, however, is oblivious to the underlying NEM structure. We propose here a suboptimal but linear complexity myopic policy for (\ref{eq:RewardMax_model2}) that allows us to generalize D-NEM to include storage operation.

The myopic scheduling algorithm is based on solving (\ref{eq:RewardMax_model2}) with a  relaxation of (\ref{eq:SoCcapacity}) that results in a closed-form solution of $\hat{\mu}^\ast$ that takes the form of a threshold-policy on the aggregated renewables $g_\Nc$, from which a generalized D-NEM can be defined naturally.  The policy $\hat{\mu}^\ast$ for the relaxed MDP (\ref{eq:RewardMax_model2}) is indeed optimal if the SoC does not exceed the storage capacity limits.
If policy $\hat{\mu}^\ast$ generates an infeasible action that violates (\ref{eq:SoCcapacity}), the proposed myopic algorithm simply projects the charging/discharging action to the feasible solution set, effectively clipping the charging/discharging amount to the upper or lower limits. See \cite{Alahmed&Tong&Zhao:23arXiv}.

\subsection{Generalized D-NEM}
\begin{figure}
    \centering
    \includegraphics[scale=0.30]{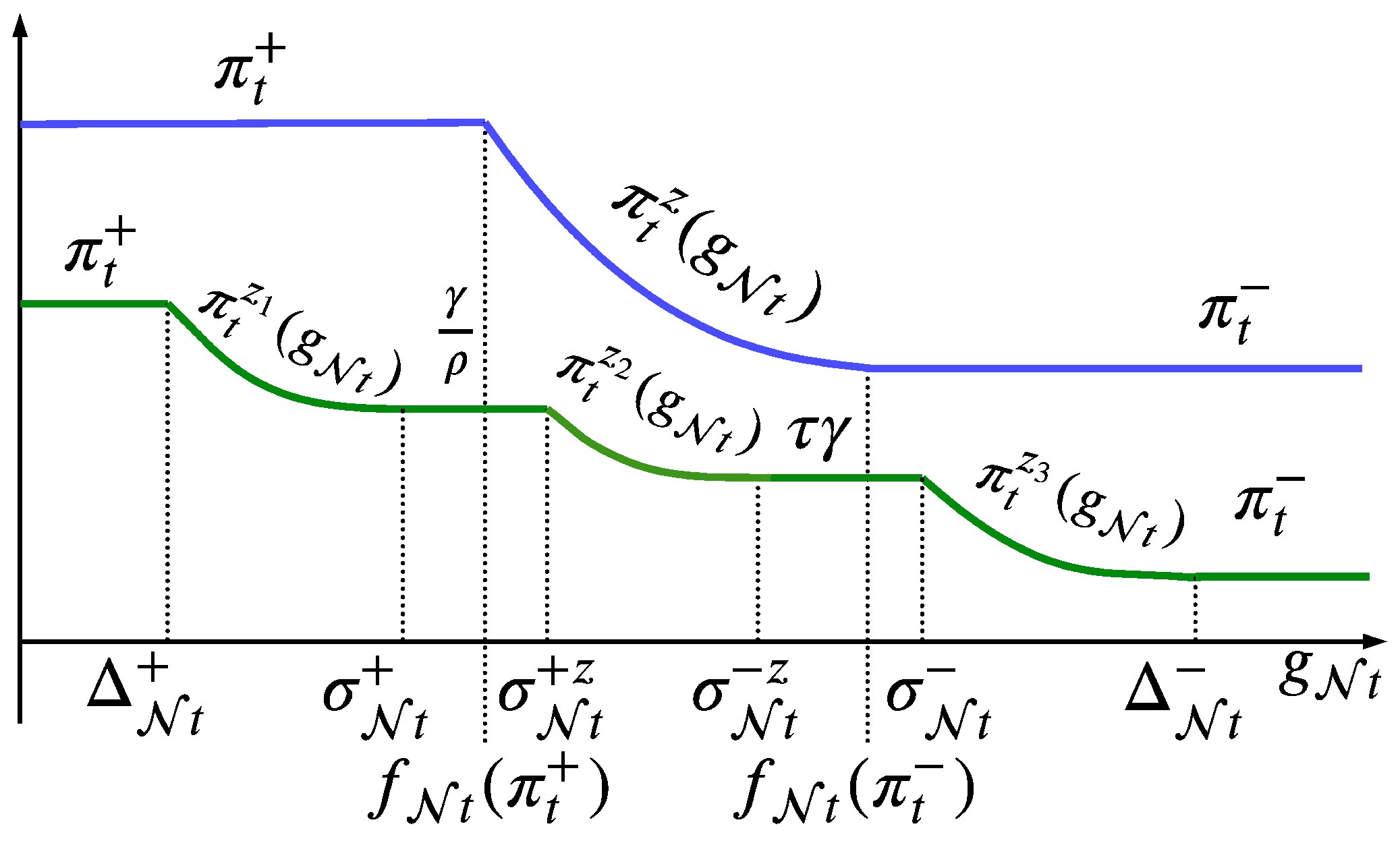}
    \vspace{-0.3cm}
    \caption{Community price under D-NEM (blue) and generalized D-NEM with BESS (green).}
    \label{fig:GenDNEM}
    \vspace{-0.7cm}
\end{figure}

Following the same bi-level setup in Sec.\ref{sec:DECWel}, the community operator devises a threshold-based pricing rule with prices announced based on $g_{\Nc t}$ for every $t$. We relegate the mathematical description of the generalized D-NEM pricing policy to Appendix D and offer instead a depiction, as shown in Fig.\ref{fig:GenDNEM}. Similar to D-NEM without BESS, the generalized D-NEM, for every $t$, passes the utility prices in the net consumption ($g_{\Nc t} \le \Delta_{\Nc t}^+$) and net production ($g_{\Nc t} \ge \Delta_{\Nc t}^-$) zones, but the size of both zones shrinks with the storage maximum energy charging/discharging ($\underline{b}_{\Nc},\overline{b}_{\Nc}$). The price when ($g_{\Nc t} \in [\Delta_{\Nc t}^+,\Delta_{\Nc t}^-]$) dynamically transitions from $\pi^+_t$ to $\pi^-_t$ as the aggregate DG $g_{\Nc t}$ increases.

Interestingly, both D-NEM and the generalized D-NEM are bounded by the DSO NEM X rates $\pi^+$ and $\pi^-$, however, the latter expands and introduces more pricing regions in the net-zero zone, \ie when ($g_{\Nc t} \in [\Delta_{\Nc t}^+,\Delta_{\Nc t}^-]$). The static prices in the net-zero zone $\gamma/\rho$ and $\tau \gamma$ when $g_{\Nc t} \in (\sigma_{\Nc t}^+,\sigma_{\Nc t}^{+z}]$ and $g_{\Nc t} \in (\sigma_{\Nc t}^{-z},\sigma_{\Nc t}^{-}]$, respectively, are announced when it is optimal for the BESS to follow $g_{\Nc t}$ and for the consumption to be fixed. The dynamic prices $\pi^{z_1}_t(g_{\Nc t})$ and $\pi^{z_3}_t(g_{\Nc t})$ are announced when maximally discharging and maximally charging the BESS, respectively, is inadequate to keep the community off-the-grid, to which the operator dynamically varies the price to incentivize the flexible demands to achieve energy balance. Lastly, the price $\pi^{z_2}_t(g_{\Nc t})$ is announced when it is more optimal to keep the storage idle and induce the demand instead, therefore, the operator does not charge/discharge the BESS, but instead changes the price to incentivize the consumption to achieve the net-zero condition.

\subsubsection{Community Operator's BESS Operation}
Based on the myopic scheduling algorithm, the output of the centrally operated BESS is a threshold policy based on $g_{\Nc t}$ \cite{Alahmed&Tong&Zhao:23arXiv}. In particular, for every $t=0,\ldots, T-1$, the BESS operation is 
        \begin{equation*}
            b^{\ast}_{\Nc t}(g_{\Nc t})=\left\{\begin{array}{ll} -\underline{\Bc}_{\Nc t}, & g_{\Nc t} \leq \sigma_{\Nc t}^+\\ g_{\Nc t}- \sigma_{\Nc t}^{+z}, & g_{\Nc t}\in (\sigma_{\Nc t}^+ , \sigma_{\Nc t}^{+z})\\ 0, & g_{\Nc t} \in [\sigma_{\Nc t}^{+z}, \sigma_{\Nc t}^{-z}]\\ g_{\Nc t}- \sigma_{\Nc t}^{-z}, & g_{\Nc t} \in (\sigma_{\Nc t}^{-z}, \sigma_{\Nc t}^{-})\\ \overline{\Bc}_{\Nc t}, & g_{\Nc t} \geq \sigma_{\Nc t}^{-},\\ \end{array}\right.
        \end{equation*}
        where
        \begin{align}
    \underline{\Bc}_{\Nc t} &:=\min\{\underline{b}_\Nc,\rho x_t\},~~~\; \overline{\Bc}_{\Nc t} :=\min\{\overline{b}_\Nc,(E-x_t)/\tau\}\nn\\
    \sigma_{\Nc t}^+ &:= f_{\Nc t}(\gamma/\rho) -\underline{\Bc}_{\Nc t},~ \sigma_{\Nc t}^{+z} :=f_{\Nc t}(\gamma/\rho)\nn\\
    \sigma_{\Nc t}^- &:= f_{\Nc t}(\tau \gamma) +\overline{\Bc}_{\Nc t},~~ \sigma_{\Nc t}^{-z}  :=f_{\Nc t}(\tau \gamma).\nn
\end{align}
Hence, for every $t=0,\ldots, T-1$, the virtual schedule of every member is $b^\ast_{it}(g_{\Nc t}):= \xi_i b^\ast_{\Nc t}(g_{\Nc t})$.

\subsection{Member's Optimal Consumption Policy}
With centralized battery operation, the operator broadcasts the community price $\pi^\ast$ along with allocated battery charging/discharging quantity $\xi_i b^\ast_{\Nc t}$.  Each individual member uses the same consumption policy $\psi^\ast$ in Sec.\ref{subsec:OptConsPol} as if there was no BESS by modifying the available $g_i$ to $\hat{g}_i=g_i - \xi_i b^\ast_{\Nc t}$.

\subsection{Properties of generalized D-NEM.}
The community pricing axioms \ref{ax:equity}--\ref{ax:ProfitNeutrality} hold for generalized D-NEM, making it a valid community pricing solution. Because it is derived from a relaxation of the centralized optimization (\ref{eq:RewardMax_model2}), it is not the price that achieves the maximum social welfare defined by (\ref{eq:RewardMax_model2}).

%% file: num_v1.tex
To show the performance of the proposed mechanism, we assumed a hypothetical energy community using one-year real data of $N=23$ residential households.\footnote{We used \href{http://www.pecanstreet.org/}{PecanStreet} 2018 data for residential users from Austin, TX.} The scheduling decisions are carried out over 24 hours horizon ($T=24$). All households have flexible loads and 18 of them have rooftop solar. The community has a BESS system with charging and discharging efficiencies $\tau=\rho = 0.95$,\footnote{This is from the roundtrip efficiency of \href{https://www.tesla.com/powerwall}{Tesla Powerwall 2}.} and power limits $\overline{b}_\mathcal{N}=\underline{b}_\mathcal{N}=23$kW. For simplicity, we assumed that community members have equivalent storage shares, \ie $\zeta_i=1/N, \forall i\in \mathcal{N}$. Throughout the results here, whenever a BESS is considered, we assume its size and initial SoC to be $E= 1.2 MWh$ and $x_0 = 0.9E$, respectively.

To model the consumption preferences of every $i\in \mathcal{N}$ household, the following utility function was adopted:
\begin{equation}\label{eq:UtilityForm}
   U_{itk}(d_{itk})=\left\{\begin{array}{ll}
\alpha_{itk} d_{itk}-\frac{1}{2}\beta_{itk} d_{itk}^2,\hspace{-0.2cm} &\hspace{-0.2cm} 0 \leq d_{itk} \leq \frac{\alpha_{ik}}{\beta_{ik}} \\
\frac{\alpha_{itk}^2}{2 \beta_{itk}},\hspace{-0.2cm} &\hspace{-0.2cm} d_{itk}>\frac{\alpha_{itk}}{\beta_{itk}},
\end{array} \right.
\end{equation}
for all $k \in \mathcal{K}, t = 0,\ldots,T-1$, where $\alpha_{itk}, \beta_{itk}$ are some utility parameters that are learned and calibrated using historical retail prices\footnote{We used \href{https://data.austintexas.gov/stories/s/EOA-C-5-a-Austin-Energy-average-annual-system-rate/t4es-hvsj/}{Data.AustinTexas.gov} historical residential rates in Austin, TX.} and consumption,\footnote{We used pre-2018 PecanStreet data for households in Austin, TX.} and by predicating an elasticity for the load type \cite{Alahmed&Tong:22EIRACM}. Two load types with two different utility functions of the form in (\ref{eq:UtilityForm}) were considered, namely 1) HVAC, and 2) other household loads.\footnote{The elasticities of HVAC and other household loads are taken from \cite{ASADINEJAD_Elasticity:18EPSR}.} Device consumption limits $\overline{\bm{d}}_i,\underline{\bm{d}}_i, \forall i\in \Nc$ were set as the maximum and minimum of the pre-2018 consumption of each load type.

The community faces the utility's NEM X tariff with a ToU rate with $\pi^+_{\mbox{\tiny Peak}}=\$0.40$/kWh and $\pi^+_{\mbox{\tiny Offpeak}}=\$0.20$/kWh as peak and offpeak prices, respectively. For the export rate $\pi^-$, we used the average 2018 real-time wholesale prices in Texas.\footnote{The data is accessible at \href{https://www.ercot.com/mktinfo/prices}{ERCOT}.} The storage salvage value rate was chosen so that $\gamma \in [\frac{1}{\tau}\max\{(\pi^-_t)\}, \rho \min\{(\pi^+_t)\}]$.

The simulation results compared the welfare, pricing, and operation of a community under D-NEM to a community under the payment rule in \cite{Chakraborty&Poolla&Varaiya:19TSG}, and to optimal standalone customers, who schedule based on \cite{Alahmed&Tong&Zhao:23arXiv} when they have storage and \cite{Alahmed&Tong:22IEEETSG} when they do not.\footnote{We assume that the customers' schedules under \cite{Chakraborty&Poolla&Varaiya:19TSG} are the same as that of {\em optimal} standalone utility customers.} In addition to comparing the market mechanisms, we evaluated the communities' performance with and without the BESS.

\subsection{Community Welfare and Surpluses With and Without BESS}
\begin{figure}
    \centering
    \includegraphics[scale=0.34]{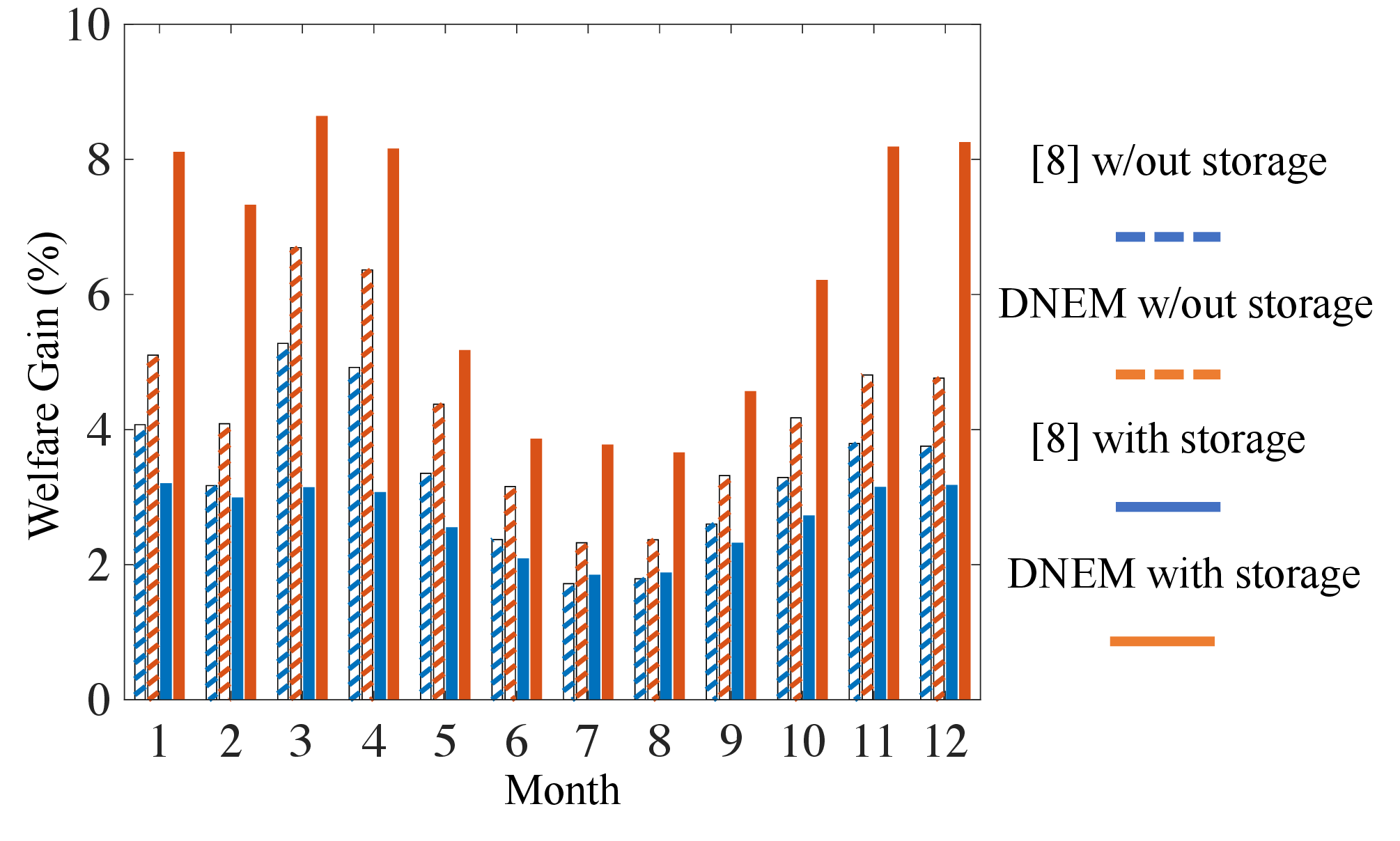}
    \vspace{-0.4cm}
    \caption{Community monthly welfare gain with and without storage.}
    \label{fig:CommWelfareNum}
\end{figure}

Fig.\ref{fig:CommWelfareNum} shows the monthly welfare gain of communities under D-NEM and \cite{Chakraborty&Poolla&Varaiya:19TSG} over standalone customers with and without BESS. In all cases, and all months, forming energy communities achieved welfare gains. The welfare gain was the highest when the aggregate net consumption was smaller (further explained in Sec.\ref{sec:NumpPrice}). Compare, for instance, the months of March--April to July--August.

In all months, the community under D-NEM, with and without BESS, achieved higher welfare gain, because, driven by price, D-NEM aligned the aggregated flexible resources with the aggregated generation, which increased renewables' self-consumption (valued at $>\pi^-$), instead of exporting to the utility (valued at $\pi^-$). Under \cite{Chakraborty&Poolla&Varaiya:19TSG}, however, every member aligned its own generation with its own flexible resources, which was sub-optimal at the aggregate level.

Looking at how each market mechanism performed with and without BESS, we note that, under D-NEM, the community's welfare gain with storage was always higher than its gain without storage. The average gain was 6.32\% with storage and 4.29\% without. This, however, was not the case under \cite{Chakraborty&Poolla&Varaiya:19TSG}, as in most months, the welfare gain of a no-storage community was higher than the welfare gain of the community with storage. This was because when customers did not have storages outside the community, they were vulnerable to the export rate $\pi^-$ more often than customers with storages, which gave the former higher incentives to join the community. Under \cite{Chakraborty&Poolla&Varaiya:19TSG}, the average gain with storage was 2.68\%, which increased to 3.34\% in the absence of storage.  

Table \ref{tab:IndMembers} offers a more specific look into Fig.\ref{fig:CommWelfareNum} by showing the average monthly member surplus gain (\%) of 13 customers over the benchmark with and without BESS and under D-NEM and \cite{Chakraborty&Poolla&Varaiya:19TSG}. Note, first, that for all 13 community members, and under both D-NEM and \cite{Chakraborty&Poolla&Varaiya:19TSG}, individual rationality was achieved. Every individual member under D-NEM achieved higher surplus gains than that under \cite{Chakraborty&Poolla&Varaiya:19TSG}. Interestingly, members under D-NEM with BESS always achieved higher gains than D-NEM without BESS. This, however, was not the case under \cite{Chakraborty&Poolla&Varaiya:19TSG} with BESS, as customers 6--7, and 11--13 had lower gains than under \cite{Chakraborty&Poolla&Varaiya:19TSG} without BESS.

\begin{table}
\centering
\caption{Average monthly member surplus gain (\%) of 13 members.}
\label{tab:IndMembers}
\vspace{-0.3cm}
\resizebox{\columnwidth}{!}{%
\begin{tabular}{@{}cccccccccccccc@{}}
\toprule \midrule
Household & 1 & 2 & 3 & 4 & 5 & 6 & 7 & 8 & 9 & 10 & 11 & 12 & 13 \\ \midrule
      \cite{Chakraborty&Poolla&Varaiya:19TSG}    &  1.3 &   2.2 &   2.2 &   1.8 &    1.4 &    8.5 &    5.4 &  1.7  &   1.7   &  1.7  &  2.9   & 4.1 &   4.6   \\ DNEM
          &   2.4  &  3.5 &   3.3 &  2.8 &   1.9 &  10.7 &   7.3  &  2.8  &  2.7 &   2.7  & 4.2   &  5.9  &  6.0      \\\midrule \cite{Chakraborty&Poolla&Varaiya:19TSG}/BESS
          &  2.6 &   2.7 &   2.7  &  2.6  &  2.1  &  4.3 &   3.6 &  2.6 &   2.3 &   2.3  &  2.9  &  3.6  &    3.4       \\ DNEM/BESS
          &  9.5  &  2.3 &  11.1 &  14.1 &   3.9 &  33.0 &  25.3 &   9.6 &  10.3 &  10.3  &  14.6  &  9.9  &  20.6     \\ 
          \midrule \bottomrule
\end{tabular}%
}
\end{table}

\subsection{Community Price and Operation}\label{sec:NumpPrice}
\begin{figure}
    \centering
    \includegraphics[scale=0.38]{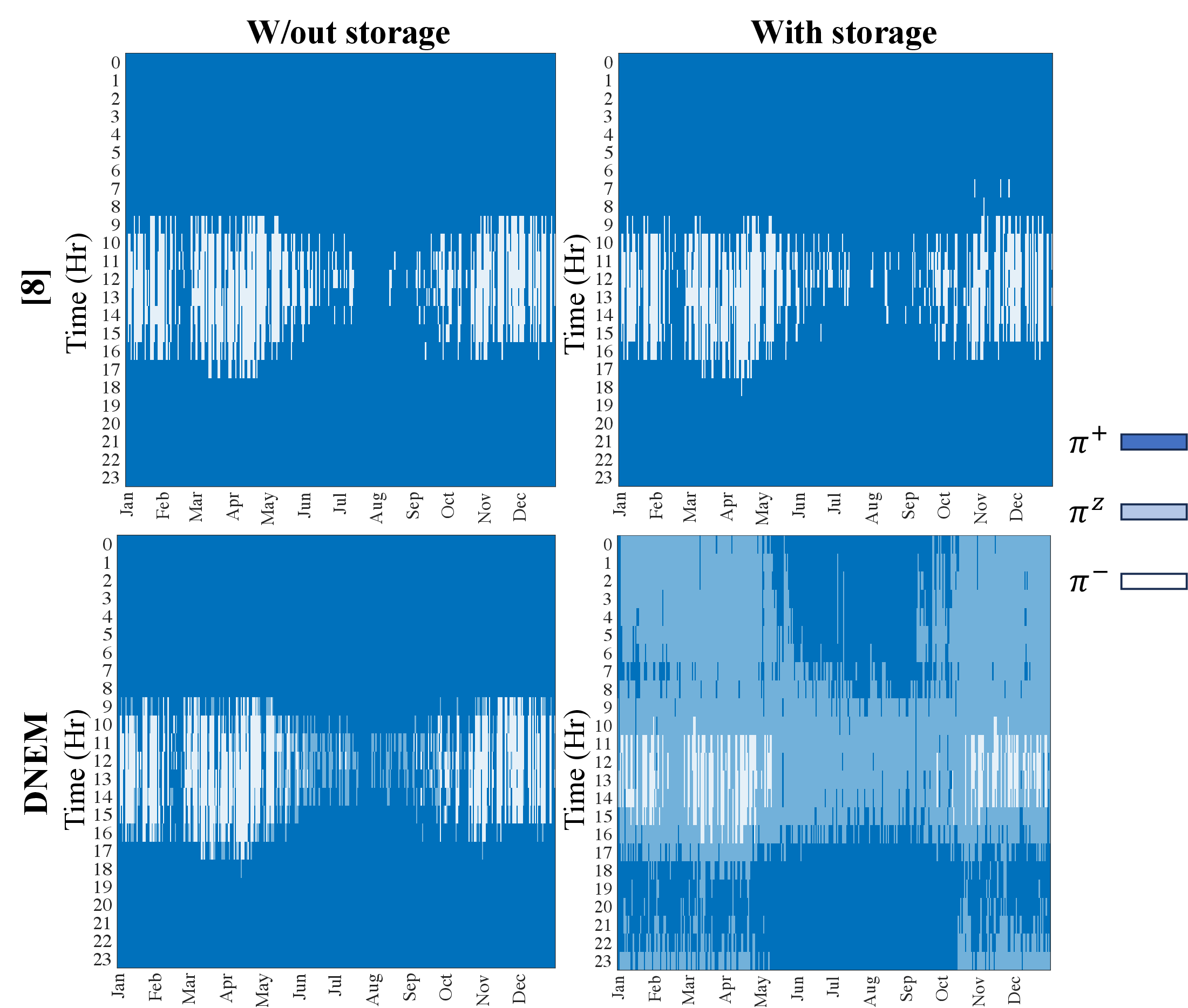}
    \vspace{-0.3cm}
    \caption{Community price with and without storage (dark blue is $\pi^+$, white is $\pi^-$ and light blue is $\pi^z(g_\mathcal{N}) \in (\pi^-,\pi^+)$).}
    \label{fig:PriceNum}
\end{figure}

\begin{figure}
    \centering
    \includegraphics[scale=0.49]{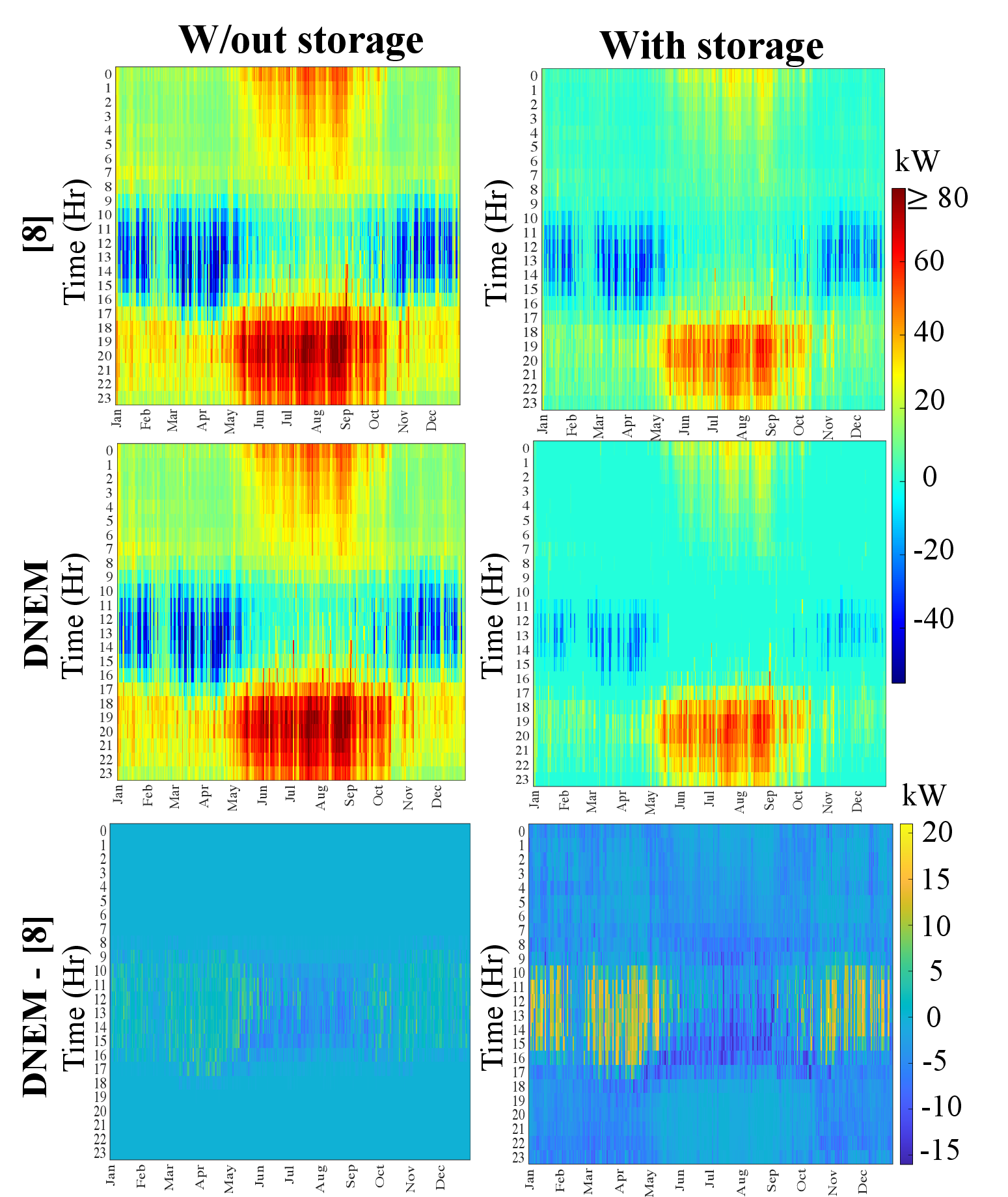}
    \vspace{-0.3cm}
    \caption{Community net consumption with and without storage.}
    \label{fig:NetConsNum}
\end{figure}

The heatmap of the community announced price (Fig.\ref{fig:PriceNum}) and the corresponding aggregate net consumption (Fig.\ref{fig:NetConsNum}) provide intuitions on why D-NEM achieved welfare optimality over \cite{Chakraborty&Poolla&Varaiya:19TSG} with and without BESS. In Fig.\ref{fig:PriceNum}, when there is no storage (left column), the price under \cite{Chakraborty&Poolla&Varaiya:19TSG} was either $\pi^+$ (dark blue) or $\pi^-$ (white) depending on the aggregate net consumption sign, whereas, under D-NEM, some of the $\pi^-$ and $\pi^+$ intervals got replaced by $\pi^z(g_\mathcal{N}) \in (\pi^-,\pi^+)$ (light blue), which drove the community's aggregate demand to equalize the aggregate generation. 
This phenomenon is reflected in Fig.\ref{fig:NetConsNum}, which shows that D-NEM managed to reduce power imports and exports through price, giving it welfare optimality over \cite{Chakraborty&Poolla&Varaiya:19TSG}. Compared to optimal standalone customers, both solar-only communities (left column in Fig.\ref{fig:NetConsNum}) reduced the midday aggregate exports, which gave them higher welfare levels (Fig.\ref{fig:CommWelfareNum}). Note that the changes to price and net consumption between D-NEM and \cite{Chakraborty&Poolla&Varaiya:19TSG} (see last row in Fig.\ref{fig:NetConsNum}) without BESS mostly occurred between 9--17, \ie during PV generation hours. Whenever the price under D-NEM and \cite{Chakraborty&Poolla&Varaiya:19TSG} are different (Fig.\ref{fig:PriceNum}), the net-consumptions are also different (Fig.\ref{fig:NetConsNum}). Without BESS, this mostly happened May--September between 9--17.

After adding a BESS (right columns of Figs.\ref{fig:PriceNum}-\ref{fig:NetConsNum}), both D-NEM and \cite{Chakraborty&Poolla&Varaiya:19TSG} further reduced power imports and exports, which increased the welfare of both communities. D-NEM was more efficient in keeping the community off-the-grid more often because the BESS and flexible consumptions were both coordinated with the aggregate renewables. As Fig.\ref{fig:PriceNum} shows, when the D-NEM community added a BESS, the net-zero zone price became more prevalent, as the storage-added-flexibility further enabled maintaining the net-zero zone.

\subsection{Effect of NEM X Rates}
\begin{figure}
\vspace{-0.45cm}
    \centering
    \includegraphics[scale=0.28]{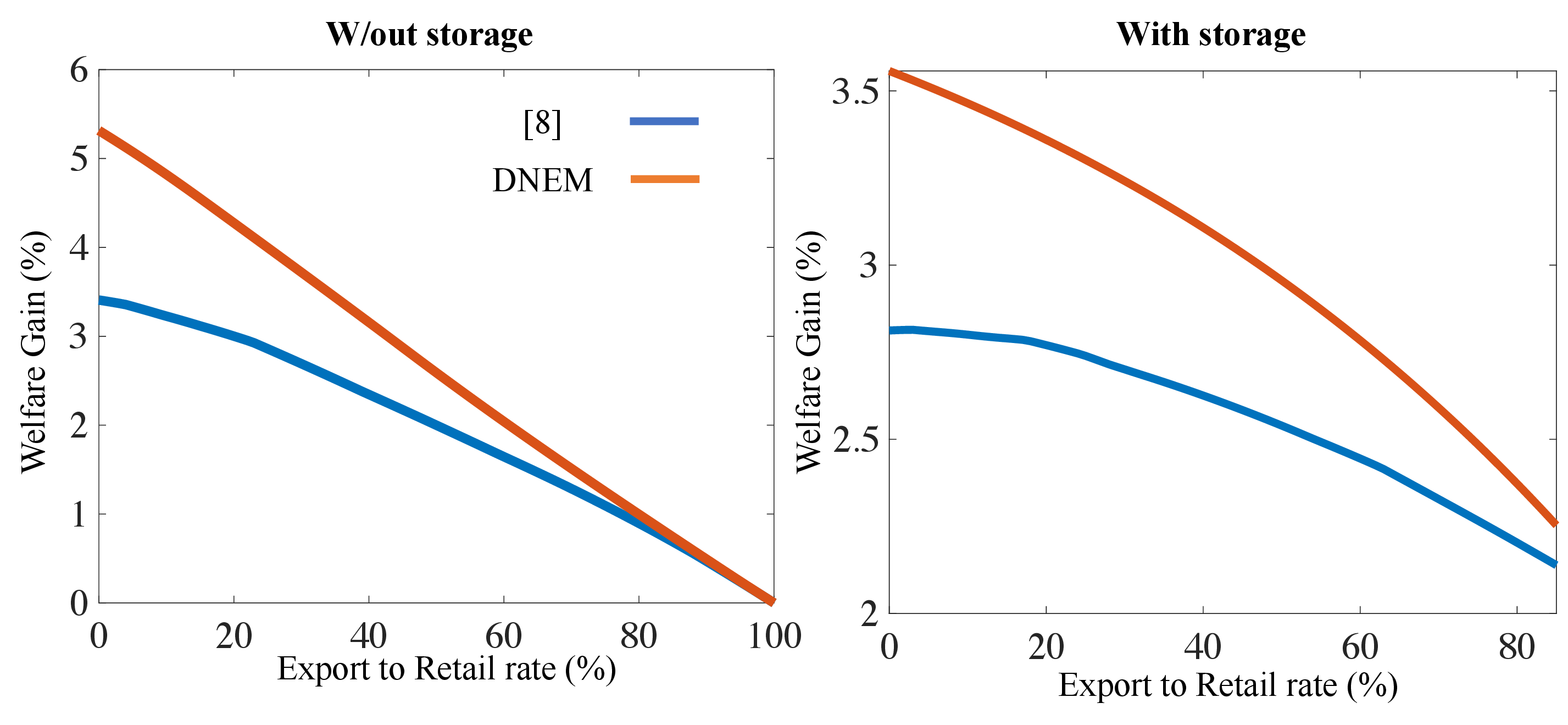}
    \vspace{-0.5cm}
    \caption{Sensitivity of welfare gain to DSO's export rate with and without storage.}
    \label{fig:SellRate}
    \vspace{-0.5cm}
\end{figure}

Figure \ref{fig:SellRate} evaluates the effect of the utility's NEM rate ratio $\pi^-/\pi^+$ on the welfare gain of communities without BESS (left) and with BESS (right), under a flat buy rate $\pi^+=\$0.4$/kWh. Three main observations are in order. First, under both \cite{Chakraborty&Poolla&Varaiya:19TSG} and D-NEM, the community welfare gain increased as $\pi^-/\pi^+$ decreased, because community members are less vulnerable to $\pi^-$ compared to their benchmark. Secondly, the welfare gain gap between \cite{Chakraborty&Poolla&Varaiya:19TSG} and D-NEM increased as $\pi^-/\pi^+$ decreased because the D-NEM community is less impacted by $\pi^-$, as through pricing, the resources are scheduled to maximize self-consumption. Lastly, the welfare gain of communities without BESS was higher with BESS at lower rate ratios, because the benchmark customers are less impacted by $\pi^-$ when they adopt BESS.

%% file: AppendixExample.tex
Consider three prosumers facing a distribution utility that bills them based on NEM X with $\pi^+ = \$0.5$/kWh, and $\pi^-=\$0.2$/kWh. The BTM renewable generations of the customers are $g_1= g_2 = 5$kW, and $g_3=0$kW. The customers have the following utility functions, and for brevity, we consider a single device case $K=1$:
\begin{align}
    U_i(d_i) &= \alpha_i \log(d_i),~ U_i(0)=0,\quad i=1,2 \nn\\
    U_i(d_i) &= \alpha_i d_i - \frac{1}{2}\beta_i d_i^2, \quad \quad \quad \quad~ i=3,\nn
\end{align}
where the utility parameters are $\alpha_1=\alpha_2=1.5$ and $\alpha_3=2, \beta_3=1$. The customers join to form an energy community $\Nc$ under D-NEM. From D-NEM, the operator can compute the thresholds as in (\ref{eq:thresholds}) to get $f_{\Nc}(\pi^+)=7.5$kW, $f_{\Nc}(\pi^-)=16.8$kW. Given that $g_\Nc=10$kW $\in [f_{\Nc}(\pi^+),f_{\Nc}(\pi^-)]$, the community is in the net-zero zone. Using (\ref{eq:NetZeroPrice}), the net-zero zone price is $\pi^{\ast}(g_\Nc)=\pi^z(g_\Nc)= \$0.36$/kWh $\in [\pi^-, \pi^+]$.

Given $\pi^z(g_\Nc)$, each prosumer solves (\ref{eq:LowerLevelOptimal}), and has the following optimal consumption $d^{\psi^\ast}_i$ and net consumption $z^{\psi^\ast}_i$:
\begin{align*}
    d^{\psi^\ast}_1 &=d^{\psi^\ast}_2= 4.18\text{kW}, \quad d^{\psi^\ast}_3 = 1.64\text{kW}, \\
    z^{\psi^\ast}_1 &=z^{\psi^\ast}_2= -0.82\text{kW}, ~ z^{\psi^\ast}_3 = 1.64\text{kW},
\end{align*}
which makes $z^{\psi^\ast}_\Nc = z^{\psi^\ast}_1 + z^{\psi^\ast}_2 + z^{\psi^\ast}_3 = 0$.
The payments and surpluses of community members are
\begin{align}
P_{1}^{\chi^\ast}&=P_{2}^{\chi^\ast}=-\$0.30, \quad P_{3}^{\chi^\ast}= \$0.59.\nn\\
    S_{1}^{\chi^\ast}&=S_{2}^{\chi^\ast}=\$2.44, \quad~~ S_{3}^{\chi^\ast}= \$1.34.\nn
\end{align}
Note that, using the derivations in \cite{Alahmed&Tong:22IEEETSG}, the payments and surpluses of customers outside the community under optimal DER scheduling decisions are:
\begin{align}
P_{1}^{{\mbox{\tiny NEM}^\ast}}&=P_{2}^{{\mbox{\tiny NEM}^\ast}}=\$0 > P_{1}^{\chi^\ast}, \quad~~ P_{3}^{{\mbox{\tiny NEM}^\ast}}= \$0.75 > P_{3}^{\chi^\ast}.\nn\\
    S_{1}^{{\mbox{\tiny NEM}^\ast}}&=S_{2}^{{\mbox{\tiny NEM}^\ast}}=\$2.41 < S_{1,2}^{\chi^\ast}, ~~ S_{3}^{{\mbox{\tiny NEM}^\ast}}= \$1.13 < S_{3}^{\chi^\ast}.\nn
\end{align}

%% file: AppendixGeneralization.tex
\subsection{Central PV}
Using the concept of VNEM \cite{NEMevolution:23NAS}, the analysis can be generalized to accommodate central PV. Households may own shares of the central community PV's output $\tilde{g}\in \mathbb{R}_{+}$. Using VNEM, the total DG generation of member
$i \in \mathcal{N}$ who owns a share $\omega_i\in \Omega=\{\omega_i\in [0,1]:\sum_{i\in \mathcal{N}} \omega_i=1\}$ of the central generation becomes $g_i + \omega_i \tilde{g}$, where $g_i$ is the BTM generation, and $\omega_i \tilde{g}$ is the output of the central generation share. Therefore, the central solar PV share's output is virtually accounted to the member as if it was a BTM resource. More importantly, prosumers can update their shares over time given their consumption and BTM DER profiles. 

\subsection{BTM BESS}
In Sec.\ref{sec:DNEM2}, we considered a centrally scheduled central BESS, however, under some assumptions, the results apply to the case when there are BTM BESS systems that are fully controllable by the community operator. Two assumptions are needed here. First, we need to ensure that all BTM BESS are scheduled using the myopic scheduling algorithm. Second, we must assume that all BTM BESS have the same charging and discharging efficiencies $\tau$, and $\rho$, respectively.

%% file: AppProofs/ProofsMain_v2.tex
\input{AppProofs/IndRationality_ProofPayment}

\input{AppProofs/MktEff_Proof}

\subsection*{Proof of Theorem \ref{thm:Equilibrium}}
\input{AppProofs/Equilibrium_proof}


\subsection*{Proof of Proposition \ref{corol:NZprice}}
\input{AppProofs/NZprice_proof_v2}

\subsection*{Proof of Theorem \ref{thm:GroupRationality}}
\input{AppProofs/GrpRationality_Proof}

%% file: AppProofs/IndRationality_ProofPayment.tex
\begin{lemma}[Cost-causation conformity of D-NEM]\label{lem:CostCausation}
D-NEM pricing policy $\chi^\ast$ conforms with the cost causation principle, \ie $\bm{P}^{\chi^\ast}\in \Ac$.
 \end{lemma}

\subsection*{Proof of Lemma \ref{lem:CostCausation}}
To prove Lemma \ref{lem:CostCausation}, we need to show that D-NEM satisfies each of Axiom \ref{ax:equity}--Axiom \ref{ax:ProfitNeutrality}.

\subsubsection*{Axiom \ref{ax:equity}}
The payments of members $i,j \in \Nc, i \neq j$ under D-NEM are given by
$$P_i^{\chi^\ast}(z_i) = \pi^\ast z_i,~~ P_j^{\chi^\ast}(z_j)=\pi^\ast z_j.$$
If $z_i=z_j$, then we have 
$$P_i^{\chi^\ast}(z_i) = \pi^\ast z_i =\pi^\ast z_j = P_j^{\chi^\ast}(z_j),$$
which is the case, because the D-NEM price $\pi^\ast$ is uniform (same for all members).
\subsubsection*{Axiom \ref{ax:monotonicity}}
For any member $i\in \Nc$, the payment under D-NEM is monotonic in $z_i$ because
$\frac{\partial P_i^{\chi^\ast}}{\partial z_i}=\pi^\ast >0.$
Additionally, the payment of every $i\in \Nc$ under D-NEM satisfies
$P_i^{\chi^\ast}(0) = \pi^\ast \cdot 0 = 0.$

\subsubsection*{Axiom \ref{ax:ProfitNeutrality}}
Given the members' optimal consumption policy $\psi^\ast$ in response to D-NEM $\chi^\ast$, the aggregate consumption of the community members is
\begin{align}
d^{\psi^\ast}_\Nc(g_\Nc)&= \sum_{i\in \Nc} \bm{1}^\top \bm{d}^{\psi^\ast}_{i}(\pi^\ast(g_{\Nc})) \nn\\&=\sum_{i\in \Nc} \bm{1}^\top \begin{cases}   \bm{d}^{\pi^+}_i&, g_\Nc<f_\Nc(\pi^+) \\ \bm{d}^{\pi^z}_i &, g_\Nc\in[f_\Nc(\pi^+),f_\Nc(\pi^-)] \\ \bm{d}^{\pi^-}_i&,
 g_\Nc>f_\Nc(\pi^-) \end{cases}\nn\\&\stackrel{\text{(\ref{eq:thresholds})}}{:=}
\begin{cases}   f_\Nc(\pi^+)&, g_\Nc<f_\Nc(\pi^+) \\ g_\Nc &, g_\Nc\in[f_\Nc(\pi^+),f_\Nc(\pi^-)] \\ f_\Nc(\pi^-)&,
 g_\Nc>f_\Nc(\pi^-) \end{cases},\nn
 \end{align}
 where 
 \begin{equation}\label{eq:Definedpi}
     \bm{d}^{y}_{i}:= \bm{d}^{\psi^\ast}_{i}(y) \stackrel{\text{(\ref{eq:optconsi})}}{:=} \max \{ \underline{\bm{d}}_{i},\min\{\bm{f}_{i}(\bm{1}\pi^\ast),\overline{\bm{d}}_{i}\}\},
 \end{equation} 
 with the $\max,\min$ operators being element-wise and $\bm{f}_i:=\bm{L}_i^{-1}$.
 Hence, the aggregate net consumption is,
 $$z^{\psi^\ast}_\Nc(g_\Nc)= \begin{cases}   f_\Nc(\pi^+) - g_\Nc&, g_\Nc<f_\Nc(\pi^+) \\ 0 &, g_\Nc\in[f_\Nc(\pi^+),f_\Nc(\pi^-)] \\ f_\Nc(\pi^-)-g_\Nc&,
 g_\Nc>f_\Nc(\pi^-) \end{cases}$$
 Therefore,
 \begin{align*}
   &\Delta P^{\chi^\ast, \mbox{\tiny NEM}}(\bm{z}^{\psi^\ast},\bm{g}) := \sum_{i\in \Nc}P^{\chi^\ast}_i(z^{\psi^\ast}_i) -  P^{\mbox{\tiny NEM}}(z^{\psi^\ast}_\Nc) =\\&
   \begin{cases}   \pi^+ z^{\psi^\ast}_\Nc(g_\Nc) &, g_\Nc<f_\Nc(\pi^+) \\ \pi^z z^{\psi^\ast}_\Nc(g_\Nc) &, g_\Nc\in[f_\Nc(\pi^+),f_\Nc(\pi^-)] \\ \pi^- z^{\psi^\ast}_\Nc(g_\Nc)&,
 g_\Nc>f_\Nc(\pi^-) \end{cases}\\ &-  \begin{cases}   \pi^+ z^{\psi^\ast}_\Nc(g_\Nc)&, z^{\psi^\ast}(g_\Nc) \geq 0  \\ \pi^- z^{\psi^\ast}_\Nc(g_\Nc)&,
 z^{\psi^\ast}(g_\Nc)<0 \end{cases}
 \end{align*}
 Note that 1) if $g_\Nc<f_\Nc(\pi^+)$ then $z^{\psi^\ast}_\Nc(g_\Nc)>0$, and $\Delta P_i^{\chi^\ast,\mbox{\tiny NEM}}=0$, 2) if $g_\Nc\in[f_\Nc(\pi^+),f_\Nc(\pi^-)]$ then $z^{\psi^\ast}_\Nc(g_\Nc)=0$ and $\Delta P_i^{\chi^\ast,\mbox{\tiny NEM}}=0$, and 3) if $g_\Nc>f_\Nc(\pi^-)$ then $z^{\psi^\ast}_\Nc(g_\Nc)<0$, and $\Delta P_i^{\chi^\ast,\mbox{\tiny NEM}}=0$, which completes the proof of profit neutrality, since $\Delta P_i^{\chi^\ast,\mbox{\tiny NEM}}=0$. 
 
 \subsubsection*{Axiom \ref{ax:rationality}}

We recall the active customer's surplus under the DSO's NEM X, derived in \cite{Alahmed&Tong:22IEEETSG}
\begin{align*}
    S^{\mbox{\tiny NEM}^\ast}_{i}(g_i)= \begin{cases}   U_{i}(\bm{d}^{\pi^+}_i)-\pi^+ (\bm{1}^\top \bm{d}^{\pi^+}_i-g_i),&\hspace{-.75em} g_i< \bm{1}^\top \bm{d}^{\pi^+}_i \\ U_{i}(\bm{d}^o_i(g_i)),&\hspace{-3.4em} g_i\in [\bm{1}^\top \bm{d}^{\pi^+}_i,\bm{1}^\top \bm{d}^{\pi^-}_i] \\ U_{i}(\bm{d}^{\pi^-}_i)-\pi^- (\bm{1}^\top \bm{d}^{\pi^-}_i-g_i),&\hspace{-.75em} g_i> \bm{1}^\top \bm{d}^{\pi^-}_i,  \end{cases}
\end{align*}
where $\bm{d}^{\pi^+}_i, \bm{d}^{\pi^-}_i$ are as defined in (\ref{eq:Definedpi}), and $\bm{d}^o_i(g_i)$ is customer's $i$ consumption in the net-zero zone, i.e., $\bm{1}^\top \bm{d}^o_i(g_i) = g_i$ when $g_i \in [\bm{1}^\top \bm{d}^{\pi^+}_i,\bm{1}^\top \bm{d}^{\pi^-}_i]$.
Under D-NEM, the surplus of every $i \in \mathcal{N}$ community member is given by:
\begin{align}
    &S^{\chi^\ast}_i(\psi^\ast(g_i),g_i)=\nn\\& \begin{cases}   U_i(\bm{d}_{i}^{\pi^+}) -\pi^+(\bm{1}^\top \bm{d}_{i}^{\pi^+}-g_i),&\hspace{-.70em} g_\mathcal{N}<f_\mathcal{N}(\pi^+) \\ U_i(\bm{d}_i^{\pi^z})-\pi^z(\bm{1}^\top \bm{d}_i^{\pi^z}-g_i), &\hspace{-.70em} g_\mathcal{N}\in[f_\mathcal{N}(\pi^+),f_\mathcal{N}(\pi^-)] \\ U_i(\bm{d}_{i}^{\pi^-})- \pi^-(\bm{1}^\top \bm{d}_{i}^{\pi^-} -g_i),& \hspace{-.70em}
 g_\mathcal{N}>f_\mathcal{N}(\pi^-). \end{cases}\nn
\end{align}
For brevity, we assume a single device case $K=1$. The additivity of $U(\cdot)$ assumption makes it straightforward to extend the proof to $K>1$. The proof assumes non-binding consumption upper and lower limits $\underline{\bm{d}}_i, \overline{\bm{d}}_i, \forall i \in \mathcal{N}$, but at the end we show how it generalizes to consumption limits inclusion. 

For every member $i \in \mathcal{N}$, the value of joining the community is $\Delta S_i^{\chi^\ast, \mbox{\tiny NEM}^\ast}(g_i):=S^{\chi^\ast}_i(\psi^\ast(g_i),g_i) - S^{\mbox{\tiny NEM}^\ast}_{i}(g_i)$, which expands to the following 9 pieces (\ref{subeq:case1})--(\ref{subeq:case9}).

\begin{subequations}\label{eq:DeltaSproof}
\begin{equation}\label{subeq:case1}
  0,~ \text{if}~~ g_\mathcal{N}<f_\mathcal{N}(\pi^+), g_i< d^{\pi^+}_i
\end{equation}    
\begin{equation}\label{subeq:case2}
  U_i(d^{\pi^+}_i)-U_i(g_i)-\pi^+\tilde{d}^{\pi^+}_i,~ \text{if}~~ g_\mathcal{N}<f_\mathcal{N}(\pi^+), g_i \in [d^{\pi^+}_i,d^{\pi^-}_i]
\end{equation}    
\begin{align}\label{subeq:case3}
  U_i(d^{\pi^+}_i)-U_i(d^{\pi^-}_i)&-\pi^+\tilde{d}^{\pi^+}_i+\pi^-\tilde{d}^{\pi^-}_i\nn\\&,~ \text{if}~~ g_\mathcal{N}<f_\mathcal{N}(\pi^+), g_i > d^{\pi^-}_i
\end{align}    
\begin{align}\label{subeq:case4}
  U_i(d^{\pi^z}_i)&-U_i(d^{\pi^+}_i)-\pi^z\tilde{d}^{\pi^z}_i+\pi^+\tilde{d}^{\pi^+}_i\nn\\&,~ \text{if}~~ g_\mathcal{N}\in [f_\mathcal{N}(\pi^+),f_\mathcal{N}(\pi^-)], g_i< d^{\pi^+}_i
\end{align}    
\begin{align}\label{subeq:case5}
  &U_i(d^{\pi^z}_i)-U_i(g_i)-\pi^z\tilde{d}^{\pi^z}_i\nn\\&,~ \text{if}~~ g_\mathcal{N}\in [f_\mathcal{N}(\pi^+),f_\mathcal{N}(\pi^-)], g_i\in [d^{\pi^+}_i,d^{\pi^-}_i]
\end{align}    
\begin{align}\label{subeq:case6}
  U_i(d^{\pi^z}_i)&-U_i(d^{\pi^-}_i)-\pi^z\tilde{d}^{\pi^z}_i+\pi^-\tilde{d}^{\pi^-}_i\nn\\&,~ \text{if}~~ g_\mathcal{N}\in [f_\mathcal{N}(\pi^+),f_\mathcal{N}(\pi^-)], g_i> d^{\pi^-}_i
\end{align}    
\begin{align}\label{subeq:case7}
  U_i(d^{\pi^-}_i)-U_i(d^{\pi^+}_i)&-\pi^-\tilde{d}^{\pi^-}_i+\pi^+\tilde{d}^{\pi^+}_i\nn\\&,~ \text{if}~~ g_\mathcal{N}>f_\mathcal{N}(\pi^-), g_i < d^{\pi^+}_i
\end{align}    
\begin{equation}\label{subeq:case8}
  U_i(d^{\pi^-}_i)-U_i(g_i)-\pi^-\tilde{d}^{\pi^-}_i,~ \text{if}~ g_\mathcal{N}>f_\mathcal{N}(\pi^-), g_i\in [d^{\pi^+}_i,d^{\pi^-}_i]
\end{equation}    
\begin{equation}\label{subeq:case9}
  0,~~ \text{if}~~ g_\mathcal{N}>f_\mathcal{N}(\pi^-), g_i> d^{\pi^-}_i
\end{equation}    
\end{subequations}

where 
$$\tilde{d}^{\pi^+}_i:=d^{\pi^+}_i-g_i,~~ \tilde{d}^{\pi^z}_i:=d^{\pi^z}_i-g_i,~~\tilde{d}^{\pi^-}_i:=d^{\pi^-}_i-g_i$$ are the renewable-adjusted optimal consumptions in the community's net-consumption, net-zero and net-production zones, respectively.

Given the concavity of $U(\cdot)$, we use the following property 
\beq \label{eq:ConcavityPropoerty}
L(x)\geq \frac{U(y)-U(x)}{y-x}\geq L(y).
\eeq
to show that $\Delta S_i^{\chi^\ast, \mbox{\tiny NEM}^\ast}\geq 0$, by proving the non-negativity of every piece in (\ref{eq:DeltaSproof}).

--  \underline{Pieces (\ref{subeq:case1}) and (\ref{subeq:case9})}: In pieces 1 and 9, the consumption of both the community member and its benchmark are $d^{\pi^+}_i$ and $d^{\pi^-}_i$, respectively. Given that they also face the same price, their surplus difference is zero.

--  \underline{Piece (\ref{subeq:case2})}: Given that $g_i>d^{\pi^+}_i$, we have:
\begin{align*}
\begin{gathered}
L_i\left(d_i^{\pi^+}\right) \geq \frac{U_i\left(g_i\right)-U_i(d_i^{\pi^+})}{g_i-d_i^{\pi^+}} \geq L_i\left(g_i\right) \\
\pi^{+}\left(d_i^{\pi^+}-g_i\right) \leq U_i(d_i^{\pi^+})-U_i\left(g_i\right) \leq L_i\left(g_i\right) \left(d_i^{\pi^+}-g_i\right) \\
0 \leq U_i(d_i^{\pi^+})-U_i\left(g_i\right)-\pi^{+}\tilde{d}^{\pi^+}_i \leq\left(L_i\left(g_i\right)-\pi^{+}\right)\tilde{d}^{\pi^+}_i
\end{gathered}
\end{align*}
where we used $L_i(d^{\pi^+}_i)=\pi^+$. Therefore, the surplus difference in piece (\ref{subeq:case2}) is non-negative.

\par --  \underline{Piece (\ref{subeq:case3})}: Given that $d^{\pi^-}_i > d^{\pi^+}_i$, we have:
\begin{equation*}
\pi^+(d^{\pi^+}_i-d^{\pi^-}_i)\leq U_i(d^{\pi^+}_i)-U_i(d^{\pi^-}_i)\leq \pi^- (d^{\pi^+}_i-d^{\pi^-}_i).
\end{equation*}

Using the lower bound $\pi^+(d^{\pi^+}_i-d^{\pi^-}_i)$, piece (\ref{subeq:case3}) becomes $(\pi^+-\pi^-)(g_i-d^{\pi^-}_i)\geq 0$, because $g_i> d^{\pi^-}_i$ and $\pi^+ \geq \pi^-$.

--  \underline{Piece (\ref{subeq:case4})}: Given that $d^{\pi^+}_i<d^{\pi^z}_i$ (from Proposition \ref{corol:NZprice}), we have:
\begin{equation*}
\pi^+ (d^{\pi^z}_i-d^{\pi^+}_i)\geq U_i(d^{\pi^z}_i)-U_i(d^{\pi^+}_i)\geq \pi^z (d^{\pi^z}_i-d^{\pi^+}_i).
\end{equation*}

Using the lower bound $\pi^z (d^{\pi^z}_i-d^{\pi^+}_i)$, we get $(\pi^+-\pi^z)(d^{\pi^+}_i-g_i)\geq0$, because $\pi^+\leq \pi^z$ (from Proposition \ref{corol:NZprice}) and $g_i<d^{\pi^+}_i$.

-- \underline{Piece(\ref{subeq:case5})}: Here, we have two possible cases.
\par Case 1: If $d^{\pi^z}_i> g_i$, we have:
\begin{equation*}
L_i(g_i) (d^{\pi^z}_i-g_i)\geq U_i(d^{\pi^z}_i)-U_i(g_i)\geq \pi^z (d^{\pi^z}_i-g_i),
\end{equation*}
which shows the non-negativity of piece (\ref{subeq:case5}).
\par Case 2: If $d^{\pi^z}_i< g_i$, then
\begin{equation*}
\pi^z (g_i-d^{\pi^z}_i)\geq U_i(g_i)-U_i(d^{\pi^z}_i)\geq L_i(g_i) (g_i-d^{\pi^z}_i).
\end{equation*}
By multiplying by ($-1$), we can see the non-negativity of (\ref{subeq:case5}).

-- \underline{Piece (\ref{subeq:case6})}: Given that $d^{\pi^z}_i<d^{\pi^-}_i$ (from Proposition \ref{corol:NZprice}):
\begin{equation*}
\pi^z (d^{\pi^z}_i-d^{\pi^-}_i)\leq U_i(d^{\pi^z}_i)-U_i(d^{\pi^-}_i)\leq \pi^- (d^{\pi^z}_i-d^{\pi^-}_i).
\end{equation*}

Using the lower bound $\pi^z (d^{\pi^z}_i-d^{\pi^-}_i)$, we get $(\pi^z-\pi^-)(g_i-d^{\pi^-}_i)\geq0$, because $\pi^z\geq\pi^-$ and $g_i>d^{\pi^-}_i$.

--  \underline{Piece (\ref{subeq:case7})}: Given $d^{\pi^-}_i>d^{\pi^+}_i$, we have:
\begin{equation*}
\pi^+(d^{\pi^-}_i-d^{\pi^+}_i)\geq U_i(d^{\pi^-}_i)-U_i(d^{\pi^+}_i)\geq \pi^- (d^{\pi^-}_i-d^{\pi^+}_i).
\end{equation*}
Using the lower bound $\pi^- (d^{\pi^-}_i-d^{\pi^+}_i)$, we get $(\pi^+-\pi^-)(d^{\pi^+}_i-g_i)\geq0$, because $\pi^+\geq \pi^-$ and $d^{\pi^+}_i>g_i$.

--  \underline{Piece (\ref{subeq:case8})}: Given $d^{\pi^-}_i>g_i$, we have:
\begin{equation*}
    L_i(g_i)(d^{\pi^-}_i-g_i)\geq U_i(d^{\pi^-}_i)-U_i(g_i)\geq \pi^- (d^{\pi^-}_i-g_i).
\end{equation*}
The inequality $U_i(d^{\pi^-}_i)-U_i(g_i)\geq \pi^- (d^{\pi^-}_i-g_i)\geq 0$ proves that piece (\ref{subeq:case8}) is non negative, because $d_i\geq g_i$.\\
Given the non-negativity of all 9 pieces (\ref{subeq:case1})--(\ref{subeq:case9}), we have $\Delta S_i^{\chi^\ast, \mbox{\tiny NEM}^\ast}\geq 0$. 

Lastly, we comment on the generalization to consumption limits. Note that if $d^{\pi^+}_i = \overline{d}_i$, then we have $d^{\pi^z}_i= d^{\pi^-}_i = \overline{d}_i$ because of the monotonicity of the inverse marginal utility $f_i$ and $\pi^+ \leq \pi^z \leq \pi^-$. Similarly, if $d^{\pi^-}_i = \underline{d}_i$, then we have $d^{\pi^z}_i= d^{\pi^+}_i = \underline{d}_i$. Given that the proof of $\Delta S_i^{\chi^\ast, \mbox{\tiny NEM}^\ast}\geq 0$ when $d^{\pi^+}_i = \overline{d}_i$ and $d^{\pi^-}_i = \underline{d}_i$ is trivial, it only remains to show the other cases. 

By proving that D-NEM satisfies Axiom \ref{ax:equity}--Axiom \ref{ax:ProfitNeutrality}, we showed that it conforms with the cost-causation principle. 
\hfill$\blacksquare$

%% file: AppProofs/MktEff_Proof.tex
 \begin{lemma}[Centralized community welfare maximization]\label{lem:CentralizedSoln}
The maximum welfare under centralized operation
\begin{align}
  W^\sharp =&\underset{(\bm{d}_i,\ldots,\bm{d}_N),\bm{z}}{\rm maximize}~ \mathbb{E}\left[\sum_{i\in \Nc} U_i(\bm{d}_i)- P^{\mbox{\tiny NEM}}(z_\Nc) \right]\nn \\& {\rm subject\; to}\hspace{0.5cm} z_\Nc = \sum_{i\in \Nc} \bm{1}^\top \bm{d}_i -g_i\label{eq:CentrWmax}\\& \hspace{2cm}
  \underline{\bm{d}}_i\preceq \bm{d}_i \preceq \overline{\bm{d}}_i,~ \forall i\in \Nc,\nn
  \end{align}
is given by
\begin{align}\label{eq:CentrWmaximum}
 &W^{\sharp}(g_\Nc)=\nonumber\\&  \begin{cases}  \sum_{i \in \Nc} U_{i}(\bm{d}^{\pi^+}_{i})-\pi^+ \left(f^{\sharp}_{\Nc}(\pi^+) - g_\Nc \right),&\hspace{-.75em} g_\Nc< f^{\sharp}_{\Nc}(\pi^+) \\ \sum_{i \in \Nc}U_{i}(\bm{d}^{\mu^\sharp}_{i}), &\hspace{-5em} g_\Nc\in[f^{\sharp}_{\Nc}(\pi^+),f^{\sharp}_{\Nc}(\pi^-)] \\ \sum_{i \in \Nc} U_{i}(\bm{d}^{\pi^-}_{i}) - \pi^-\left(f^{\sharp}_{\Nc}(\pi^-) - g_\Nc \right), &\hspace{-.75em} g_\Nc> f^{\sharp}_{\Nc}(\pi^-), \end{cases}
\end{align}   
where $\bm{d}^{y}_{i} \in \mathbb{R}^K_+$ and $f^{\sharp}_{\Nc}(y) \in \mathbb{R}_+$ are, respectively, defined by
\begin{align}
\bm{d}^{y}_{i} := \max \{\underline{\bm{d}}_{i},\min\{\bm{f}_i(\bm{1}y),\bar{\bm{d}}_i\}\}, f^{\sharp}_{\Nc}(y):= \sum_{i\in \Nc} \bm{1}^\top \bm{d}^{y}_{i}\nn,
\end{align}
with the $\max,\min$ operators being element-wise and $\bm{f}_i:=\bm{L}_i^{-1}$ is the inverse marginal utility vector. The price $\mu^\sharp(g_\Nc) \in [\pi^-, \pi^+]$ is the Lagrangian multiplier solving:
\begin{equation}\label{eq:OptZeroZone_Model2}
 \sum_{i \in \Nc} \sum_{k \in \mathcal{K}} \max\{\underline{d}_{ik}, \min\{f_{ik}(\mu),\bar{d}_{ik}\}\} = g_\Nc.
\end{equation}
\end{lemma}

\subsection*{Proof of Lemma \ref{lem:CentralizedSoln}}
The program in (\ref{eq:CentrWmax}) is a generalization to the standalone consumer decision problem under the DSO's NEM X regime in \cite{Alahmed&Tong:22IEEETSG} with an additional dimension representing the $N$ members. Therefore, the optimal consumption $\bm{d}^{\sharp}_i \in \mathbb{R}_+^K$ and net consumption $z^{\sharp}_i \in \mathbb{R}$ of every $i \in \Nc$ member both obey a two threshold-policy, respectively as
\begin{align}
    \bm{d}^{\sharp}_i (g_\Nc)&= \begin{cases}  \bm{d}_i^{\pi^+},&\hspace{-.75em} g_\Nc< f^{\sharp}_{\Nc}(\pi^+) \\ \bm{d}_i^{\mu^\sharp}, &\hspace{-.75em} g_\Nc\in[f^{\sharp}_{\Nc}(\pi^+),f^{\sharp}_{\Nc}(\pi^-)] \\ \bm{d}_i^{\pi^-}, &\hspace{-.75em} g_\Nc>f^{\sharp}_{\Nc}(\pi^-),\end{cases} \label{eq:Optmemberd} \\  z^{\sharp}_i (g_\Nc)&= \bm{1}^\top \bm{d}^{\sharp}_i (g_\Nc) - g_{\Nc},
\end{align}
where $\bm{d}_i^{y}$ is as defined in Lemma \ref{lem:CentralizedSoln}. The social welfare under optimal decisions is
\begin{align*}
W^{\sharp}(g_\Nc) &= \sum_{i\in \Nc} U_i(\bm{d}^\sharp_i(g_\Nc)) -P^{\mbox{\tiny NEM}}(z^\sharp_\Nc),
\end{align*}
where $z^\sharp_\Nc(g_\Nc):=\sum_{i\in \Nc} \bm{1}^\top \bm{d}^{\sharp}_i (g_\Nc) - g_\Nc$. By substituting (\ref{eq:Optmemberd}) in the equation above, one should get the maximum social welfare in (\ref{eq:CentrWmaximum}). Note that the expectation is irrelevant here because the solution optimizes for every possible realization, therefore it optimizes the expected value too. \hfill$\blacksquare$

%% file: AppProofs/Equilibrium_proof.tex
The stochastic bi-level optimization in Sec.\ref{sec:DECWel} can be compactly formulated as 
\begin{align}
 \underset{P(\cdot), \{\bm{d}_i\}_{i=1}^N}{\operatorname{maximize}}& \Bigg(W^{\chi_\psi} = \mbbE\Big[\sum_{i\in \Nc} U_i(\bm{d}_i^\psi) - P^{\mbox{\tiny NEM}}(z_{\Nc})\Big]\Bigg)\nn\\
 \text{subject to}&~~~ z_{\Nc} = \sum_{i\in \Nc} \bm{1}^\top \bm{d}_i^\psi- g_i\nn\\
 &~~~ \text{ for all } i=1, \ldots, N\nn\\
 &~~~ \bm{d}^{\psi}_i := \underset{\bm{d}_i\in \mathbb{R}_+^K}{\operatorname{argmax}}~~ U_i(\bm{d}_i) - P^\chi_i(z_i)\nn\\
 &\hspace{2.1cm} \underline{\bm{d}}_i \preceq \bm{d}_i \preceq \overline{\bm{d}}_i,\nn
\end{align}
where we also removed the constraint $P(\cdot) \in \Ac$ from the upper level problem, hence we solve an upper bound that does not constrain the payment rule to the set of cost-causation conforming payment rules $\Ac$. Next, we reformulate the program above by (1) replacing the lower-level optimization problem by its KKT conditions (mathematical program with equilibrium constraints \cite{Luo&Pang&Ralph:96Cambridge}), and (2) using $P^\chi_i(z_i):= \pi \cdot z_i, \forall i \in \Nc$ from Axioms \ref{ax:equity}--\ref{ax:monotonicity}, as
\begin{align}
 \underset{\pi, \{\overline{\bm{\lambda}}\}_{i=1}^N,\{\underline{\bm{\lambda}}\}_{i=1}^N}{\operatorname{maximize}}& \Bigg(W^{\chi_\psi} = \mbbE\Big[\sum_{i\in \Nc} U_i(\bm{d}_i^\psi) - P^{\mbox{\tiny NEM}}(z_{\Nc})\Big]\Bigg)\nn\\
 \text{subject to}&~~~ z_{\Nc} = \sum_{i\in \Nc} \sum_{k\in \Kc} d_{ik}^\psi- g_i,\nn\\
 &~~~ \text{ for all } i=1, \ldots, N, k=1,\ldots, K\nn\\
 &~~~ L_{ik}(d_{ik}^\psi)-\pi+ \overline{\lambda}_{ik} -  \underline{\lambda}_{ik}=0 \nn\\
 &\hspace{1.6cm} \overline{d}_{ik} - d_{ik}^\psi \geq 0 \perp \overline{\lambda}_{ik} \geq 0 \nn\\
 &\hspace{1.6cm} d_{ik}^\psi-\underline{d}_{ik} \geq  0 \perp \underline{\lambda}_{ik} \geq 0, \nn
\end{align}
where $x \perp y$ means that $x$ and $y$ are perpendicular. By implementing the complementarity conditions above into the constraint $L_{ik}(d_{ik}^\psi)-\pi+ \overline{\lambda}_{ik} -  \underline{\lambda}_{ik}=0$, we have
\begin{align}
 \underset{\pi}{\operatorname{maximize}}& \Bigg(W^{\chi_\psi} = \mbbE\Big[\sum_{i\in \Nc} U_i(\bm{d}_i^\psi) - P^{\mbox{\tiny NEM}}(z_{\Nc})\Big]\Bigg)\nn\\
 \text{subject to}&~~~ z_{\Nc} = \sum_{i\in \Nc} \sum_{k\in \Kc} d_{ik}^\psi- g_i,\nn\\
 &~~~ \text{ for all } i=1, \ldots, N, k=1,\ldots, K\nn\\
 &~~~ d_{ik}^\psi= \max\{\underline{d}_{ik},\min\{f_{ik}(\pi),\overline{d}_{ik}\} \}. \nn
\end{align}
 Now, note that if $\pi^\ast$ is found, we have equilibrium. Under NEM X tariff (\ref{eq:Pbenchmark}), if $z_{\Nc}>0$ then $P^{\mbox{\tiny NEM}}(z_{\Nc}) = \pi^+ z_{\Nc} > 0$, and to achieve profit-neutrality (Axiom \ref{ax:ProfitNeutrality},$\sum_{i\in \Nc} P^\chi_i(z_i)=P^{\mbox{\tiny NEM}}(z_{\Nc})$), we must have $\pi^\ast = \pi^+$. This will result in $d_{ik}^\psi= \max\{\underline{d}_{ik},\min\{f_{ik}(\pi^+),\overline{d}_{ik}\} \}, \forall k\in \Kc, \forall i \in \Nc$, which when aggregated gives $\sum_{i\in \Nc} \sum_{k\in \Kc} d_{ik}^\psi = f_{\Nc}(\pi^+) $. Hence, the community price is $\pi^\ast = \pi^+$ only if $g_{\Nc}< f_{\Nc}(\pi^+)$, which makes $z_{\Nc}>0$.

 Similarly, if $z_{\Nc}<0$ then $P^{\mbox{\tiny NEM}}(z_{\Nc}) = \pi^- z_{\Nc} < 0$, and to achieve profit-neutrality, we must have $\pi^\ast = \pi^-$. This will result in $d_{ik}^\psi= \max\{\underline{d}_{ik},\min\{f_{ik}(\pi^-),\overline{d}_{ik}\} \}, \forall k\in \Kc, \forall i \in \Nc$, which when aggregated gives $\sum_{i\in \Nc} \sum_{k\in \Kc} d_{ik}^\psi = f_{\Nc}(\pi^-) $. The community price is $\pi^\ast = \pi^-$ only if $g_{\Nc}> f_{\Nc}(\pi^-)$, which makes $z_{\Nc}<0$.

 Lastly, if $z_{\Nc}=0$ then $P^{\mbox{\tiny NEM}}(z_{\Nc}) = 0$. Because $z_{\Nc}=0 \rightarrow \sum_{i\in \Nc} \sum_{k\in \Kc} d_{ik}^\psi = g_{\Nc}$, the community price $\pi^\ast(g_{\Nc})$ is the one that solves
 $$\sum_{i\in \Nc} \sum_{k\in \Kc} \max\{\underline{d}_{ik},\min\{f_{ik}(\pi),\overline{d}_{ik}\} \}= g_{\Nc},$$
 which is equivalent to (\ref{eq:NetZeroPrice}), and hence $\pi^\ast(g_{\Nc}) = \pi^z(g_{\Nc})$, whenever $g_{\Nc}\in [f_{\Nc}(\pi^+),f_{\Nc}(\pi^-)]$.

 Thus, in summary, the equilibrium community price is 
 $$\pi^\ast(g_\Nc)=\begin{cases}   \pi^+&, g_\Nc<f_\Nc(\pi^+) \\ \pi^z(g_\Nc) &, g_\Nc\in[f_\Nc(\pi^+),f_\Nc(\pi^-)] \\ \pi^-&,
 g_\Nc>f_\Nc(\pi^-) \end{cases},$$
 and the payment to every community member is $P_i^{\chi^\ast}(z_i) = \pi^\ast(g_\Nc) \cdot z_i$, which is equivalent to D-NEM in (\ref{eq:PricingMechanism}). By leveraging Lemma \ref{lem:CostCausation}, we can see that $\bm{P}^{\chi^\ast} \in \Ac$ conforms with the cost-causation principle.

To show that the equilibrium achieves the highest community welfare, we leverage Lemma \ref{lem:CentralizedSoln}. Note that under D-NEM, the member's optimal consumption policy
$$
d_{ik}^{\psi^\ast}(\pi^\ast) = \max \{ \underline{d}_{ik},\min\{f_{ik}(\pi^\ast),\bar{d}_{ik}\}\}, \forall k \in \Kc, \forall i\in \Nc,
$$
results in the following surplus function
$$S^{\chi^\ast}_i(\psi^\ast(g_i),g_i) = U_i(\bm{d}_{i}^{\psi^\ast}(\pi^\ast)) - P^{\chi^\ast}_i(z_{i}^{\psi^\ast}(\pi^\ast)),$$
where $z_{i}^{\psi^\ast}(\pi^\ast) = \bm{1}^\top \bm{d}_{i}^{\psi^\ast}(\pi^\ast) - g_i$. By aggregating the surplus of members under their optimal consumption policy
\begin{align}
    &\hspace{-0.1cm}\sum_{i\in \Nc}\hspace{-0.08cm}S^{\chi^\ast}_i(\psi^\ast(g_i),g_i) \hspace{-0.1cm}=\hspace{-0.18cm} \sum_{i\in \Nc} \Big(U_i(\bm{d}_{i}^{\psi^\ast}(\pi^\ast)) - P^{\chi^\ast}_i(z_{i}^{\psi^\ast}(\pi^\ast))\Big)\nn\\\stackrel{\text{(A)}}{=}& \hspace{-0.08cm}\sum_{i\in \Nc} \begin{cases} U_i(\bm{d}_{i}^{\psi^\ast}(\pi^+)) -  \pi^+ z_{i}^{\psi^\ast}(\pi^+)&, g_\Nc<f_\Nc(\pi^+) \\ U_i(\bm{d}_{i}^{\psi^\ast}(\pi^z)) -  \pi^z z_{i}^{\psi^\ast}(\pi^z) &\hspace{-0.56cm}, g_\Nc\in[f_\Nc(\pi^+),f_\Nc(\pi^-)] \\ U_i(\bm{d}_{i}^{\psi^\ast}(\pi^-)) -  \pi^- z_{i}^{\psi^\ast}(\pi^-)&, g_\Nc>f_\Nc(\pi^-)\end{cases}\nn\\\stackrel{\text{(\ref{eq:NetZeroPrice})}}{=}&  \begin{cases} \sum_{i\in \Nc} U_i(\bm{d}_{i}^{\psi^\ast}(\pi^+)) -  \pi^+ z_{\Nc}^{\psi^\ast}(\pi^+)&, g_\Nc<f_\Nc(\pi^+) \\ \sum_{i\in \Nc} U_i(\bm{d}_{i}^{\psi^\ast}(\pi^z))  &\hspace{-1.3cm}, g_\Nc\in[f_\Nc(\pi^+),f_\Nc(\pi^-)] \\ \sum_{i\in \Nc} U_i(\bm{d}_{i}^{\psi^\ast}(\pi^-)) -  \pi^- z_{\Nc}^{\psi^\ast}(\pi^-)&,
 g_\Nc>f_\Nc(\pi^-)\end{cases} \nn\\=:& W^{\chi^\ast,\psi^\ast} \geq \sum_{i\in \Nc} S^{\chi}_i(\psi^\ast(g_i),g_i), \nn
\end{align}
where (A) is done by leveraging D-NEM threshold structure, and the expectation is dropped because the solution is for every possible realization, therefore it includes the expected value too. From Lemma \ref{lem:CentralizedSoln}, we can see that $W^{\chi^\ast,\psi^\ast} = W^\sharp$, and that $\bm{d}^{\psi^\ast}_i = \bm{d}^{\sharp}_i, z^{\psi^\ast}_i = z^{\sharp}_i$, for all $i \in \Nc$, which proves that the community welfare under the equilibrium solution $W^{\chi^\ast,\psi^\ast}$ is the maximum community welfare under centralized operation $W^\sharp$. 
\hfill$\blacksquare$

%% file: AppProofs/NZprice_proof_v2.tex
We showed in Lemma \ref{lem:CentralizedSoln} that the D-NEM price $\pi^z(g_\mathcal{N}):=\mu^\ast(g_\mathcal{N})$ when $g_\mathcal{N} \in [f_{\Nc}(\pi^+),f_{\Nc}(\pi^-)]$ (from Theorem \ref{thm:Equilibrium}), $f_{\Nc}(\pi^+) = f_{\Nc}^\sharp(\pi^+), f_{\Nc}(\pi^-) = f_{\Nc}^\sharp(\pi^-)$) is the solution of:
 \begin{equation*}
 \sum_{i \in \mathcal{N}} \sum_{k \in \mathcal{K}} \max\{\underline{d}_{ik}, \min\{f_{ik}(\mu),\bar{d}_{ik}\}\} = g_\mathcal{N}.
\end{equation*}
By letting
\begin{equation*}
 F(x):= \sum_{i \in \mathcal{N}} \sum_{k \in \mathcal{K}} \max\{\underline{d}_{ik}, \min\{f_{ik}(\mu),\bar{d}_{ik}\}\} - g_\mathcal{N},
\end{equation*}
one could see that $F(\cdot)$ is a continuous and monotonically decreasing function of $g_\mathcal{N}$. When $g_\mathcal{N} \in [f_{\Nc}(\pi^+),f_{\Nc}(\pi^-)]$, we have
\[ F(\pi^+)\leq 0,\quad \quad F(\pi^-)\geq 0.\]
Therefore, there must exists $\mu \in [\pi^-,\pi^+], \text{s.t. }F(\mu)=0$. Furthermore, the continuity and monotonicity of $F(\cdot)$ in $g_\mathcal{N}$ implies that $\mu$ is a continuous and monotonically decreasing function of $g_\mathcal{N}$. \hfill$\blacksquare$

%% file: AppProofs/GrpRationality_Proof.tex
Similar to the proof of Axiom \ref{ax:rationality} in Lemma \ref{lem:CostCausation}, we will assume for brevity that $K=1$, and non-binding consumption upper and lower limits $\underline{\bm{d}}_i, \overline{\bm{d}}_i, \forall i \in \mathcal{N}$. Generalization to include consumption limits is shown at the end of the proof of Axiom \ref{ax:rationality} in Lemma \ref{lem:CostCausation}.

Under D-NEM, the surplus of every member under coalition $\Sc$ is
\begin{align*}
&S^{\chi^\ast}_{i,\Sc}(\psi^\ast(g_i),g_i)=\nn\\ &\begin{cases}U_i(d_i^{\pi^+})-\pi^+(d_i^{\pi^+}-g_i) & , g_{\Sc}\leq f_{\Sc}(\pi^+) \\ U_i(d_i^{\pi^z(g_\Sc)})-\pi^z(g_{\Sc})(d_i^{\pi^z(g_\Sc)}\hspace{-0.3cm}-g_i) & \hspace{-0.32cm} , g_{\Sc}\hspace{-0.1cm}\in \hspace{-0.1cm}[f_{\Sc}(\pi^+),f_{\Sc}(\pi^-)]  \\ U_i(d_i^{\pi^-})-\pi^-(d_i^{\pi^-}-g_i) & , g_{\Sc}\geq f_{\Sc}(\pi^-).\end{cases}\nn
\end{align*}
The surplus under the coalition $\Hc \subseteq \Sc$ and D-NEM is:
\begin{align*}
&S^{\chi^\ast}_{i,\Hc}(\psi^\ast(g_i),g_i)=\nn\\&\begin{cases}U_i(d_i^{\pi^+})-\pi^+(d_i^{\pi^+}-g_i) & , g_{\Hc}\leq f_{\Hc}(\pi^+) \\ U_i(d_i^{\pi^z(g_\Hc)})-\pi^z(g_{\Hc})(d_i^{\pi^z(g_\Hc)}\hspace{-0.3cm}-g_i) &\hspace{-0.32cm} , g_{\Hc}\hspace{-0.1cm}\in \hspace{-0.1cm}[f_{\Hc}(\pi^+),f_{\Hc}(\pi^-)]  \\ U_i(d_i^{\pi^-})-\pi^-(d_i^{\pi^-}-g_i) & , g_{\Hc}\geq f_{\Hc}(\pi^-).\end{cases}
\end{align*}

We will consider all there cases of the mother coalition $\Sc$: 1) $g_{\Sc}\leq f_{\Sc}(\pi^+)$, 2) $g_{\Sc}\in [f_{\Sc}(\pi^+),f_{\Sc}(\pi^-)]$, and 3) $g_{\Sc}\geq f_{\Sc}(\pi^-)$.
\begin{enumsteps}[wide, labelwidth=!, labelindent=0pt]
\item $g_{\Sc}\leq f_{\Sc}(\pi^+)$\\
Here, we have $$\sum_{i\in \Hc} S^{\chi^\ast}_{i,\Sc}=\sum_{i\in \Hc} \left( U_i(d_i^{\pi^+})-\pi^+(d_i^{\pi^+}-g_i)\right).$$ Now, we need to show that $\sum_{i\in \Hc} S^{\chi^\ast}_{i,\Sc}\geq \sum_{i \in \Hc}S^{\chi^\ast}_{i,\Hc}$. We first write:
\begin{align*}
&\sum_{i \in \Hc} S^{\chi^\ast}_{i,\Hc}=\nn\\&\begin{cases} \sum_{i \in \Hc} \left(U_i(d_i^{\pi^+})-\pi^+(d_i^{\pi^+}-g_i)\right) & , g_{\Hc}\leq f_{\Hc}(\pi^+) \\ \sum_{i \in \Hc} \left(U_i(d_i^{\pi^z(g_\Hc)})\right) &\hspace{-1cm} , g_{\Hc}\in [f_{\Hc}(\pi^+),f_{\Hc}(\pi^-)]  \\ \sum_{i \in \Hc} \left(U_i(d_i^{\pi^-})-\pi^-(d_i^{\pi^-}-g_i)\right) & , g_{\Hc}\geq f_{\Hc}(\pi^-),\end{cases}\nn
\end{align*}
and note that we have 3 sub-cases.
\begin{enumsteps}[wide, labelwidth=!, labelindent=0pt]
\item $g_{\Hc}\leq f_{\Hc}(\pi^+)$\\
If $g_{\Hc}\leq f_{\Hc}(\pi^+)$, then $\sum_{i\in \Hc} S^{\chi^\ast}_{i,\Sc}= \sum_{i \in \Hc}S^{\chi^\ast}_{i,\Hc}$. 
\item $g_{\Hc}\geq f_{\Hc}(\pi^-)$\\
We want to show that
\begin{align*}
\sum_{i \in \Hc} \Big(U_i(d_i^{\pi^-})-\pi^-(d_i^{\pi^-}&-g_i)\Big)\\& \leq \sum_{i \in \Hc} \Big(U_i(d_i^{\pi^+})-\pi^+(d_i^{\pi^+}-g_i)\Big)
\end{align*}
From the concavity of $U(\cdot)$, we have:
$$
\pi^+ \geq \frac{U_i(d^{\pi^-}_i)-U_i(d^{\pi^+}_i)}{d^{\pi^-}_i-d^{\pi^+}_i}\geq \pi^-
$$
where we used $L_i(d^{\pi^+}_i) = \pi^+$ and $L_i(d^{\pi^-}_i) = \pi^-$.
By introducing $\sum_{i \in \Hc}$, to the inequalities above, 
we use the upper bound $\sum_{i \in \Hc}\pi^+(d^{\pi^-}_i-d^{\pi^+}_i)$ to get
$$
(\pi^+-\pi^-)(f_{\Hc}(\pi^-)-g_\Hc)\leq 0
$$
because $\pi^+\geq \pi^-$ and $g_{\Hc}\geq f_{\Hc}(\pi^-)$.
\item $g_{\Hc}\in [f_{\Hc}(\pi^+),f_{\Hc}(\pi^-)]$\\
Similarly, for this case, we have:

\begin{align*}
\pi^+(g_\Hc-f_{\Hc}(\pi^+)) &\geq \sum_{i \in \Hc} \Big(U_i(d^z_i)-U_i(d^{\pi^+}_i)\Big)\\&\geq \pi^z(g_\Hc-f_{\Hc}(\pi^+))
\end{align*}

by taking $\sum_{i \in \Hc}S^{\chi^\ast}_{i,\Hc}-\sum_{i\in \Hc} S^{\chi^\ast}_{i,\Sc}$, we have from the inequality above:

$$
\left(U_i(d^z_i)-U_i(d^{\pi^+}_i)\right) - \pi^+(g_\Hc-f_{\Hc}(\pi^+))\leq 0
$$
\end{enumsteps}
\item $g_{\Sc}\geq f_{\Sc}(\pi^-)$\\
Here, we have $$\sum_{i\in \Hc} S^{\chi^\ast}_{i,\Sc}=\sum_{i\in \Hc} \left( U_i(d_i^{\pi^-})-\pi^-(d_i^{\pi^-}-g_i)\right).$$

\begin{enumsteps}[wide, labelwidth=!, labelindent=0pt]
\item $g_{\Hc}\geq f_{\Hc}(\pi^-)$\\ Then $\sum_{i\in \Hc} S^{\chi^\ast}_{i,\Sc}= \sum_{i \in \Hc}S^{\chi^\ast}_{i,\Hc}$.
\item $g_{\Hc}\leq f_{\Hc}(\pi^+)$\\
Here we use: 
$$
\pi^+ \geq \frac{U_i(d^{\pi^-}_i)-U_i(d^{\pi^+}_i)}{d^{\pi^-}_i-d^{\pi^+}_i}\geq \pi^-
$$
to show that 
\begin{align*}
\sum_{i \in \Hc} \Big(U_i(d_i^{\pi^-})-\pi^-(&d_i^{\pi^-}-g_i)\Big)\\& \geq \sum_{i \in \Hc} \Big(U_i(d_i^{\pi^+})-\pi^+(d_i^{\pi^+}-g_i)\Big),
\end{align*}
we use the lower bound $\pi^- (f_{\Hc}(\pi^-)-f_{\Hc}(\pi^+))$, and get $$\sum_{i\in \Hc} S^{\chi^\ast}_{i,\Sc}- \sum_{i \in \Hc}S^{\chi^\ast}_{i,\Hc}=(\pi^+-\pi^-)(f_{\Hc}(\pi^+)-g_\Hc)\geq 0.$$
\item $g_{\Hc}\in [f_{\Hc}(\pi^+),f_{\Hc}(\pi^-)]$\\
Similarly, for this case, we have:
\begin{align*}
\pi^z(f_{\Hc}(\pi^-)-g_\Hc) &\geq \sum_{i \in \Hc} \Big(U_i(d^{\pi^-}_i)-U_i(d^z_i)\Big)\\&\geq \pi^-(f_{\Hc}(\pi^-)-g_\Hc).
\end{align*}
 Hence, by computing the surplus difference, we get:
 \begin{align*}
 \sum_{i\in \Hc} S^{\chi^\ast}_{i,\Sc}- \sum_{i \in \Hc}S^{\chi^\ast}_{i,\Hc}&=\sum_{i \in \Hc} \left(U_i(d^{\pi^-}_i)-U_i(d^z_i)\right)\nn\\&-\pi^-(f_{\Hc}(\pi^-)-g_\Hc)\geq 0.\nn
 \end{align*}
 \end{enumsteps}
\item $g_{\Sc}\in [f_{\Sc}(\pi^+), f_{\Sc}(\pi^-)]$\\
Here, we have $\sum_{i\in \Hc} S^{\chi^\ast}_{i,\Sc}=\sum_{i\in \Hc} U_i(d_i^z(g_\Sc))$.

We have 3 cases.
\begin{enumsteps}[wide, labelwidth=!, labelindent=0pt]
\item $g_{\Hc}\leq f_{\Hc}(\pi^+)$\\
Here we have: 
\begin{align}
\sum_{i\in \Hc} S^{\chi^\ast}_{i,\Sc}- \sum_{i \in \Hc}S^{\chi^\ast}_{i,\Hc}&=\sum_{i \in \Hc} \left(U_i(d_i^z(g_\Sc))-U_i(d^{\pi^+}_i)\right)\nn\\&-\pi^z(g_{\Sc})(\sum_{i\in \Hc}d_i^{\pi^z(g_\Sc)}-g_\Hc)\nn\\&+\pi^+(f_{\Hc}(\pi^+)-g_\Hc).\nn
\end{align}
We use the lower bound in:
\begin{align}
\pi^+(\sum_{i\in \Hc}d_i^{\pi^z(g_\Sc)}-f_{\Hc}(\pi^+))&\geq \sum_{i\in \Hc}(U_i(d^z_i(g_\Sc))-U_i(d^{\pi^+}_i))\nn\\&\geq \pi^z(g_\Sc) (\sum_{i\in \Hc}d_i^{\pi^z(g_\Sc)}-f_{\Hc}(\pi^+))\nn
\end{align}
and substitute to get
$$
 \sum_{i\in \Hc} S^{\chi^\ast}_{i,\Sc}- \sum_{i \in \Hc}S^{\chi^\ast}_{i,\Hc}=(\pi^+-\pi^z(g_\Sc))(f_{\Hc}(\pi^+)-g_\Hc)\geq0.
$$
\item $g_{\Hc}\geq f_{\Hc}(\pi^-)$\\
Here we have: 
\begin{align}
\sum_{i\in \Hc} S^{\chi^\ast}_{i,\Sc}- \sum_{i \in \Hc}S^{\chi^\ast}_{i,\Hc}&=\sum_{i \in \Hc} \left(U_i(d_i^z(g_\Sc))-U_i(d^{\pi^-}_i)\right)\nn\\&-\pi^z(g_{\Sc})(\sum_{i\in \Hc}d_i^{\pi^z(g_\Sc)}-g_\Hc)\nn\\&+\pi^-(f_{\Hc}(\pi^-)-g_\Hc)\nn
\end{align}
We use the lower bound in:
\begin{align}
\pi^z(g_\Sc)(\sum_{i\in \Hc}d_i^{\pi^z(g_\Sc)}\hspace{-0.18cm}-f_{\Hc}(\pi^-))&\leq \hspace{-0.1cm}\sum_{i\in \Hc}(U_i(d^z_i(g_\Sc))-U_i(d^{\pi^-}_i))\nn\\& \leq \pi^- (\sum_{i\in \Hc}d_i^{\pi^z(g_\Sc)}-f_{\Hc}(\pi^-))\nn
\end{align}
and substitute to get
$$
 \sum_{i\in \Hc} S^{\chi^\ast}_{i,\Sc}- \sum_{i \in \Hc}S^{\chi^\ast}_{i,\Hc}=(\pi^z(g_\Sc)-\pi^-)(g_\Hc-f_{\Hc}(\pi^-))\geq0.
$$
because $g_{\Hc}\geq f_{\Hc}(\pi^-)$ and $\pi^z(g_\Sc)\geq\pi^-$.
\item $g_{\Hc}\in [f_{\Hc}(\pi^+),f_{\Hc}(\pi^-)]$\\
Here, we have 
\begin{align}
\sum_{i\in \Hc} S^{\chi^\ast}_{i,\Sc}- \sum_{i \in \Hc}S^{\chi^\ast}_{i,\Hc}&=\sum_{i \in \Hc} \left(U_i(d_i^z(g_\Sc))-U_i(d_i^z(g_\Hc))\right)\nn\\&-\pi^z(g_{\Sc})(\sum_{i\in \Hc}d_i^{\pi^z(g_\Sc)}-g_\Hc)\nn
\end{align}
If $d_i^z(g_\Sc)\leq d_i^z(g_\Hc)$, then we know from the monotonicity of the marginal utility that $\pi^z(g_\Sc)\geq \pi^z(g_\Hc)$.
Using the lower bound of:
\begin{align}
&\sum_{i \in \Hc} \pi^z(g_\Sc)(d_i^z(g_\Sc)-d_i^z(g_\Hc)) \leq\nn\\& \sum_{i \in \Hc} \left(U_i(d_i^z(g_\Sc))-U_i(d_i^z(g_\Hc))\right)\leq \nn\\& \sum_{i \in \Hc}\pi^z(g_\Hc)(d_i^z(g_\Sc)-d_i^z(g_\Hc))\nn
\end{align}
we get:
$$
\sum_{i\in \Hc} S^{\chi^\ast}_{i,\Sc}- \sum_{i \in \Hc}S^{\chi^\ast}_{i,\Hc}=\pi^z(g_\Sc)(g_\Hc-g_\Hc)=0
$$
Alternatively, if $d_i^z(g_\Sc)\geq d_i^z(g_\Hc)$, then we know that $\pi^z(g_\Sc)\leq \pi^z(g_\Hc)$. Using the lower bound of:
\begin{align}
&\sum_{i \in \Hc} \pi^z(g_\Sc)(d_i^z(g_\Sc)-d_i^z(g_\Hc)) \leq\nn\\& \sum_{i \in \Hc} \left(U_i(d_i^z(g_\Sc))-U_i(d_i^z(g_\Hc))\right)\leq\nn\\& \sum_{i \in \Hc}\pi^z(g_\Hc)(d_i^z(g_\Sc)-d_i^z(g_\Hc))\nn
\end{align}
we get:
$$
\sum_{i\in \Hc} S^{\chi^\ast}_{i,\Sc}- \sum_{i \in \Hc}S^{\chi^\ast}_{i,\Hc}=\pi^z(g_\Sc)(g_\Hc-g_\Hc)=0
$$ 
\end{enumsteps}
\end{enumsteps}
Hence, $\sum_{i\in \Hc} S^{\chi^\ast}_{i,\Sc}(\psi^\ast(g_i),g_i)\geq \sum_{i \in \Hc}S^{\chi^\ast}_{i,\Hc}(\psi^\ast(g_i),g_i)$ is proved.

Lastly, we comment on the generalization to consumption limits. Note that if $d^{\pi^+}_i = \overline{d}_i$, then we have $d^z_i= d^{\pi^-}_i = \overline{d}_i$ because of the monotonicity of the inverse marginal utility $f_i$ and $\pi^+ \leq \pi^z \leq \pi^-$. Similarly, if $d^{\pi^-}_i = \underline{d}_i$, then we have $d^z_i= d^{\pi^+}_i = \underline{d}_i$.
\hfill$\blacksquare$

%% file: AppendixGenDNEM.tex
When a central BESS is added, and given the aggregate renewable DG $g_{\Nc}$, the operator announces the community pricing rule according to the generalized Dynamic NEM policy shown below
\begin{policy1*}\label{alloc:GeneralizeddDNEM}
Under the generalized D-NEM, for every $t\in [0, T-1]$, the community pricing rule is threshold-based $\chi^\ast_t: \bm{g}_t \mapsto P^{\chi^\ast}_t(z_t)=\pi^\ast_t(g_{\Nc t})\cdot z_t,$ with
\begin{equation}
\pi_t^{\ast}(g_{\mathcal{N}t})= \left\{\begin{array}{ll} \pi^+_t, & g_{\mathcal{N}t} < \Delta_{\mathcal{N}t}^+\\ \pi^{z_1}_t(g_{\mathcal{N}t}), & g_{\mathcal{N}t} \in [\Delta_{\mathcal{N}t}^+,\sigma_{\mathcal{N}t}^{+}]\\ \gamma/\rho, & g_{\mathcal{N}t}\in (\sigma_{\mathcal{N}t}^+, \sigma_{\mathcal{N}t}^{+z})\\ \pi^{z_2}_t(g_{\mathcal{N}t}), & g_{\mathcal{N}t} \in [\sigma_{\mathcal{N}t}^{+z}, \sigma_{\mathcal{N}t}^{-z}]\\ \gamma\tau, & g_{\mathcal{N}t}\in (\sigma_{\mathcal{N}t}^{-z},\sigma_{\mathcal{N}t}^{-})\\ \pi^{z_3}_t(g_{\mathcal{N}t}), & g_{\mathcal{N}t}\in [\sigma_{\mathcal{N}t}^{-},\Delta_{\mathcal{N}t}^{-}]\\ \pi^-_t, & g_{\mathcal{N}t} > \Delta_{\mathcal{N}t}^-,\\ \end{array}\right.
\end{equation}
where the thresholds are computed, for $t\in[0, T-1]$, as:
  \begin{align}
      \begin{array}{ll} 
      \Delta_{\mathcal{N}t}^+:= f_{\mathcal{N}t}(\pi_t^+)-\underline{\Bc}_{\Nc t},~~&\Delta_{\mathcal{N}t}^-:= f_{\mathcal{N}t}(\pi_t^-) + \overline{\Bc}_{\Nc t},\nonumber\\ \sigma_{\mathcal{N}t}^+:= f_{\mathcal{N}t}(\gamma/\rho)- \underline{\Bc}_{\Nc t}, &\sigma_{\mathcal{N}t}^-:= f_{\mathcal{N}t}(\tau\gamma) + \overline{\Bc}_{\Nc t},\\ \sigma_{\mathcal{N}t}^{+z}:= f_{\mathcal{N}t}(\gamma/\rho), &\sigma_{\mathcal{N}t}^{-z}:= f_{\mathcal{N}t}(\tau\gamma),\nonumber\\ \end{array}
  \end{align}
  and $\underline{\Bc}_{\Nc t}:=\min\{\underline{b}_\Nc,\rho x_t\},~ \overline{\Bc}_{\Nc t} :=\min\{\overline{b}_\Nc,(E-x_t)/\tau\}\nn$. The prices $\pi^+_t,\gamma/\rho,\gamma \tau,\pi^-_t$ are constants, and the dynamic prices $\pi^{z_1}_t(g_{\mathcal{N}t}),\pi^{z_2}_t(g_{\mathcal{N}t}),\pi^{z_3}_t(g_{\mathcal{N}t})$ are the solutions of
 \begin{align}
 f_{\mathcal{N}t}(\mu_t^{z_1})&=g_{\mathcal{N}t}+\underline{\Bc}_{\Nc t}\nonumber\\ f_{\mathcal{N}t}(\mu_t^{z_2})&=g_{\mathcal{N}t}\nonumber\\ f_{\mathcal{N}t}(\mu_t^{z_3})&=g_{\mathcal{N}t}-\overline{\Bc}_{\Nc t},\nonumber
  \end{align} 
respectively.
\end{policy1*}

%% file: Alahmed_Tong_ECstorage_v3.bbl
\begin{thebibliography}{10}
\providecommand{\url}[1]{#1}
\csname url@samestyle\endcsname
\providecommand{\newblock}{\relax}
\providecommand{\bibinfo}[2]{#2}
\providecommand{\BIBentrySTDinterwordspacing}{\spaceskip=0pt\relax}
\providecommand{\BIBentryALTinterwordstretchfactor}{4}
\providecommand{\BIBentryALTinterwordspacing}{\spaceskip=\fontdimen2\font plus
\BIBentryALTinterwordstretchfactor\fontdimen3\font minus \fontdimen4\font\relax}
\providecommand{\BIBforeignlanguage}[2]{{%
\expandafter\ifx\csname l@#1\endcsname\relax
\typeout{** WARNING: IEEEtran.bst: No hyphenation pattern has been}%
\typeout{** loaded for the language `#1'. Using the pattern for}%
\typeout{** the default language instead.}%
\else
\language=\csname l@#1\endcsname
\fi
#2}}
\providecommand{\BIBdecl}{\relax}
\BIBdecl

\bibitem{parag&sovacool:16Nature}
Y.~Parag and B.~K. Sovacool, ``Electricity market design for the prosumer era,'' \emph{Nature Energy}, vol.~1, no.~4, Mar 2016.

\bibitem{Lezama&Soares&Hernandez:19TPS}
F.~Lezama, J.~Soares, P.~Hernandez-Leal, M.~Kaisers, T.~Pinto, and Z.~Vale, ``Local energy markets: Paving the path toward fully transactive energy systems,'' \emph{IEEE Trans. on Power Systems}, no.~5, 2019.

\bibitem{CitizenEC:EuroCommWebsite}
\BIBentryALTinterwordspacing
``Energy communities, {European Commission}.'' [Online]. Available: \url{https://energy.ec.europa.eu/topics/markets-and-consumers/energy-communities_en}
\BIBentrySTDinterwordspacing

\bibitem{Cohen&Kietzmann:14OrgEnvJSTOR}
\BIBentryALTinterwordspacing
B.~Cohen and J.~Kietzmann, ``Ride on! mobility business models for the sharing economy,'' \emph{Organization \& Environment}, vol.~27, no.~3, pp. 279--296, 2014. [Online]. Available: \url{http://www.jstor.org/stable/26164716}
\BIBentrySTDinterwordspacing

\bibitem{NEMevolution:23NAS}
{National Academies of Sciences, Engineering, and Medicine}, \emph{The Role of Net Metering in the Evolving Electricity System}.\hskip 1em plus 0.5em minus 0.4em\relax The National Academies Press, 2023.

\bibitem{Abada&Ehrenmann&Lambin:20EP}
I.~Abada, A.~Ehrenmann, and X.~Lambin, ``Unintended consequences: The snowball effect of energy communities,'' \emph{Energy Policy}, 2020.

\bibitem{Yang&Guoqiang&Spanos:21TSG}
Y.~Yang, G.~Hu, and C.~J. Spanos, ``Optimal sharing and fair cost allocation of community energy storage,'' \emph{IEEE Transactions on Smart Grid}, vol.~12, no.~5, Sep. 2021.

\bibitem{Chakraborty&Poolla&Varaiya:19TSG}
P.~{Chakraborty}, E.~{Baeyens}, P.~P. {Khargonekar}, K.~{Poolla}, and P.~{Varaiya}, ``Analysis of solar energy aggregation under various billing mechanisms,'' \emph{IEEE Transactions on Smart Grid}, vol.~10, no.~4, 2019.

\bibitem{Chis&Koivunen:19TSG}
A.~Chis and V.~Koivunen, ``{Coalitional game-based cost optimization of energy portfolio in smart grid communities},'' \emph{IEEE Transactions on Smart Grid}, vol.~10, no.~2, 2019.

\bibitem{Han&Morstyn&McCulloch:19TPS}
L.~Han, T.~Morstyn, and M.~McCulloch, ``Incentivizing prosumer coalitions with energy management using cooperative game theory,'' \emph{IEEE Transactions on Power Systems}, vol.~34, no.~1, 2019.

\bibitem{Fleischhacker&Auer&Lettner&Botterud:19TSG}
A.~Fleischhacker, H.~Auer, G.~Lettner, and A.~Botterud, ``Sharing solar {PV} and energy storage in apartment buildings: Resource allocation and pricing,'' \emph{IEEE Transactions on Smart Grid}, vol.~10, no.~4, 2019.

\bibitem{Cui&Wang&Yan&Shi&Xiao:21TSG}
S.~Cui, Y.-W. Wang, Y.~Shi, and J.-W. Xiao, ``Community energy cooperation with the presence of cheating behaviors,'' \emph{IEEE Transactions on Smart Grid}, vol.~12, no.~1, 2021.

\bibitem{Guerreroetal:20RSER}
J.~Guerrero, D.~Gebbran, S.~Mhanna, A.~C. Chapman, and G.~Verbič, ``Towards a transactive energy system for integration of distributed energy resources: Home energy management, distributed optimal power flow, and peer-to-peer energy trading,'' \emph{Renewable and Sustainable Energy Reviews}, vol. 132, 2020.

\bibitem{Hao&Wu&Lian&Yang:188TSG}
H.~Hao, D.~Wu, J.~Lian, and T.~Yang, ``{Optimal Coordination of Building Loads and Energy Storage for Power Grid and End User Services},'' \emph{IEEE Transactions on Smart Grid}, vol.~9, no.~5, 2018.

\bibitem{Kalathil&Wu&Poolla&Varaiya:19TSG}
D.~Kalathil, C.~Wu, K.~Poolla, and P.~Varaiya, ``{The Sharing Economy for the Electricity Storage},'' \emph{IEEE Transactions on Smart Grid}, vol.~10, no.~1, pp. 556--567, 2019.

\bibitem{Tsaousoglou&Giraldo&Paterakis:22RSER}
G.~Tsaousoglou, J.~S. Giraldo, and N.~G. Paterakis, ``Market mechanisms for local electricity markets: A review of models, solution concepts and algorithmic techniques,'' \emph{Renewable and Sustainable Energy Reviews}, vol. 156, 2022.

\bibitem{Morstyn&Teytelboym&Mcculloch:19TSG}
T.~Morstyn, A.~Teytelboym, and M.~D. Mcculloch, ``Bilateral contract networks for peer-to-peer energy trading,'' \emph{IEEE Transactions on Smart Grid}, vol.~10, no.~2, 2019.

\bibitem{Sorin&Bobo&Pinson:19TPS}
E.~Sorin, L.~Bobo, and P.~Pinson, ``Consensus-based approach to peer-to-peer electricity markets with product differentiation,'' \emph{IEEE Transactions on Power Systems}, vol.~34, no.~2, 2019.

\bibitem{Prete&Hobbs:16AE}
C.~{Lo Prete} and B.~F. Hobbs, ``A cooperative game theoretic analysis of incentives for microgrids in regulated electricity markets,'' \emph{Applied Energy}, vol. 169, 2016.

\bibitem{Cui&Wang&Li&Xiao:20}
S.~Cui, Y.-W. Wang, C.~Li, and J.-W. Xiao, ``{Prosumer community: A risk aversion energy sharing model},'' \emph{IEEE Transactions on Sustainable Energy}, vol.~11, no.~2, 2020.

\bibitem{Celik&Roche&Bouquain&Miraoui:18TSG}
B.~Celik, R.~Roche, D.~Bouquain, and A.~Miraoui, ``Decentralized neighborhood energy management with coordinated smart home energy sharing,'' \emph{IEEE Transactions on Smart Grid}, vol.~9, no.~6, 2018.

\bibitem{Chakraborty&Baeyens&Poolla&Khargonekar&Varaiya:19TSG}
P.~Chakraborty, E.~Baeyens, K.~Poolla, P.~P. Khargonekar, and P.~Varaiya, ``{Sharing Storage in a Smart Grid: A Coalitional Game Approach},'' \emph{IEEE Transactions on Smart Grid}, vol.~10, no.~4, 2019.

\bibitem{Shapley&Shubik:73RAND}
L.~S. Shapley and M.~Shubik, \emph{Game Theory in Economics: Chapter 6, Characteristic Function, Core, and Stable Set}.\hskip 1em plus 0.5em minus 0.4em\relax RAND, 1973.

\bibitem{bonbright1988principles}
J.~Bonbright, A.~Danielsen, D.~Kamerschen, and J.~Legler, \emph{Principles of Public Utility Rates}.\hskip 1em plus 0.5em minus 0.4em\relax Public Utilities Reports, 1988.

\bibitem{Alahmed&Tong:23PESGM}
A.~S. Alahmed and L.~Tong, ``Achieving social optimality for energy communities via dynamic {NEM} pricing,'' in \emph{2023 IEEE Power and Energy Society General Meeting (PESGM)}, 2023, pp. 1--5.

\bibitem{Alahmed&Tong:22IEEETSG}
------, ``On net energy metering {X}: Optimal prosumer decisions, social welfare, and cross-subsidies,'' \emph{IEEE Transactions on Smart Grid}, vol.~14, no.~02, 2023.

\bibitem{Alahmed&Tong&Zhao:23arXiv}
\BIBentryALTinterwordspacing
A.~S. Alahmed, L.~Tong, and Q.~Zhao, ``Co-optimizing distributed energy resources in linear complexity under net energy metering,'' 2023. [Online]. Available: \url{https://arxiv.org/abs/2208.09781#}
\BIBentrySTDinterwordspacing

\bibitem{Liu&Ochoa&Wong&Theunissen:22TSG}
M.~Z. Liu, L.~F. Ochoa, P.~K.~C. Wong, and J.~Theunissen, ``Using {OPF}-based operating envelopes to facilitate residential der services,'' \emph{IEEE Transactions on Smart Grid}, vol.~13, no.~6, pp. 4494--4504, 2022.

\bibitem{Alahmed&Cavraro&Bernstein&Tong:23AllertonArXiv}
A.~S. Alahmed, G.~Cavraro, A.~Bernstein, and L.~Tong, ``Operating-envelopes-aware decentralized welfare maximization for energy communities,'' in \emph{2023 59th Annual Allerton Conference on Communication, Control, and Computing (Allerton)}, 2023, pp. 1--8.

\bibitem{Alahmed&Tong:22EIRACM}
A.~S. Alahmed and L.~Tong, ``Integrating distributed energy resources: Optimal prosumer decisions and impacts of net metering tariffs,'' \emph{SIGENERGY Energy Inform. Rev.}, vol.~2, no.~2, Aug. 2022.

\bibitem{Hu&Zhang:21TSG}
L.~He and J.~Zhang, ``A community sharing market with {PV} and energy storage: An adaptive bidding-based double-side auction mechanism,'' \emph{IEEE Transactions on Smart Grid}, vol.~12, no.~3, 2021.

\bibitem{Driessen&Funaki:1991}
T.~S. Driessen and Y.~Funaki, ``Coincidence of and collinearity between game theoretic solutions,'' \emph{OR Spectrum}, vol.~13, no.~1, Mar. 1991.

\bibitem{Chakraborty&Baeyens&Khargonekar:18TPS}
P.~Chakraborty, E.~Baeyens, and P.~P. Khargonekar, ``Cost causation based allocations of costs for market integration of renewable energy,'' \emph{IEEE Transactions on on Power Systems}, vol.~33, no.~1, 2018.

\bibitem{Chen&Wang&Heo&Kishore:13TSG}
C.~Chen, J.~Wang, Y.~Heo, and S.~Kishore, ``{MPC}-based appliance scheduling for residential building energy management controller,'' \emph{IEEE Transactions on Smart Grid}, vol.~4, no.~3, pp. 1401--1410, 2013.

\bibitem{ASADINEJAD_Elasticity:18EPSR}
A.~Asadinejad, A.~Rahimpour, K.~Tomsovic, H.~Qi, and C.~fei Chen, ``Evaluation of residential customer elasticity for incentive based demand response programs,'' \emph{Electric Power Systems Research}, 2018.

\bibitem{Luo&Pang&Ralph:96Cambridge}
Z.-Q. Luo, J.-S. Pang, and D.~Ralph, \emph{Mathematical Programs with Equilibrium Constraints}.\hskip 1em plus 0.5em minus 0.4em\relax Cambridge University Press, 1996.

\end{thebibliography}
